\documentclass[12pt]{article}
\usepackage{amssymb,amsmath,epsfig,graphicx,cite,color}

\numberwithin{equation}{section}
\numberwithin{table}{section}

\DeclareMathOperator \nor { \begin{picture}(2.5,9)(0,0) \put(-1.5,-2){$\stackrel{\circ}{\scriptstyle \circ}$} \end{picture} }
\DeclareMathOperator \nol {  \begin{picture}(4,9)(0,0) \put(-.5,-2){$\stackrel{\circ}{\scriptstyle \circ}$}\end{picture} }

\begin{document}

\baselineskip=16pt

\begin{titlepage}
{}~ \hfill\vbox{\hbox{UK/08-10} }\break
\vskip 4.1cm

\centerline{\Large \bf Singular gauge transformations in string field theory} \vspace*{2.0ex}
\vspace*{8.0ex}

\centerline{\large \rm Ian Ellwood}

\vspace*{8.0ex}

\centerline{\large \it Department of Physics and Astronomy, } 
\centerline{\large \it University of Kentucky, Lexington, KY 40506, USA} 
\vspace*{2.0ex}
\centerline{E-mail: {\tt iellwood@pa.uky.edu}}

\vspace*{6.0ex}

\vspace*{6.0ex}

\centerline{\bf Abstract}
\bigskip

\noindent
We show that the existence of a tachyon vacuum puts tight restrictions on the form of solutions to the equations
of motion of string field theory.  In particular, we demonstrate that every solution can be written as a -- formal --  gauge transformation of the
tachyon vacuum.  In order for a solution to be non-trivial, this gauge transformation must be singular and we argue that this will happen when the gauge transformation annihilates a projector of the star-algebra.  We comment on possible applications of the formalism to finding new solutions.
\end{titlepage}
\tableofcontents


\section{Introduction}
\label{s:intro}

An unexpected discovery from the ``analytic revolution'' \cite{Schnabl:2005gv,Okawa:2006vm,Fuchs:2006hw,Ellwood:2006ba,Erler:2006hw,Okawa:2006sn,Rastelli:2006ap,Schnabl:2007az,Kiermaier:2007ba,Okawa:2007ri,Erler:2007rh,Fuchs:2007gw,Fuchs:2007yy,Kiermaier:2007vu,Rastelli:2007gg,Ellwood:2007xr,Ellwood:2008jh,Kiermaier:2008qu,Fuchs:2008zx,Fuchs:2008cc,Aref'eva:2009ac} in our understanding of string field theory \cite{Witten:1985cc}, is that the tachyon vacuum solution found by Schnabl \cite{Schnabl:2005gv} can be written formally as a gauge transformation of the perturbative vacuum,
\begin{equation}\label{PsiAsPureGauge}
   \Psi \simeq V^{-1} Q_B V  \ .
\end{equation}
This form is, to say the least, disconcerting, as $\Psi$ is obviously not pure-gauge; it has both a different energy and a different BRST cohomology than the perturbative vacuum. While there has been work on understanding how to regulate $\Psi$ so that it gives the correct answers for the tachyon vacuum (as opposed to the perturbative vacuum) \cite{Schnabl:2005gv,Fuchs:2006hw,Okawa:2006vm,Erler:2006hw,Erler:2007xt,Aref'eva:2009ac}, there has been little motivation for why it should, nonetheless, take a pure-gauge form.  Indeed, nothing in the original derivation of the solution was based on a pure-gauge ansatz.   The solution was only put into pure-gauge form later by Okawa \cite{Okawa:2006vm} (see, however, \cite{Schnabl:2000cp}).

Given how unexpected it is, we might wonder if the form (\ref{PsiAsPureGauge}) is telling us something about the structure of SFT solutions. It fact, as we will show, it is one example of a general rule:  {\em every} solution to the equations of motion can be written -- formally -- as a  {\em gauge transformation of the  tachyon vacuum}.  This rule relies on only one assumption: that the BRST operator around the tachyon vacuum has no cohomology at any ghostnumber so that one can construct a homotopy field $A$ as was done in \cite{Ellwood:2006ba}.  

If we let $\mathcal{Q}$ be the BRST operator around the tachyon vacuum and $\Phi$ a solution to the equations of motion, $\mathcal{Q}\Phi + \Phi * \Phi = 0$, it is possible to find an analytic expression for a  $U(\Phi,A)$ such that $\Phi = U^{-1} \mathcal{Q} U$.  True solutions (as opposed to gauge rotations of the tachyon vacuum) are then characterized by the way in which the operator $U(\Phi,A)$ fails to be an actual gauge transformation.  For the known solutions, this happens in one particular way: there exists a projector, which we call the characteristic projector, which is in the right kernel of $U$,
\begin{equation}
  UP =0 \ .
\end{equation}
Because of this,  $U^{-1}P $ diverges, which we take as evidence that $U^{-1}$ is singular\footnote{In general, in string field theory we have no rigorous definition of when a string field is well-behaved.  It may be, because projectors are outside the space of nice string fields, that $U^{-1} P$ diverging is not enough to make $U^{-1}$ itself singular.  Lacking a more rigorous approach, we stick with our intuition that $U^{-1}$ is singular and leave aside the intriguing possibility that $U$ should be though of, instead, as a ``large'' gauge transformation.}.  In general, if one is looking for gauge transformations satisfying the reality condition $U^{\ddagger} = U^{-1}$, both $U$ and $U^{-1}$ will be singular in this fashion.

The existence of the characteristic projector $P$ is thus required in a non-trivial string field theory solution: it is an obstruction to showing that the solution is gauge trivial.  We will find in each of the known cases that $P$ is a rank one sliver-like projector with unmodified boundary conditions near its endpoints and new boundary conditions at its midpoint.  It can be computed by taking a limit
\begin{equation}
   P = \lim_{N \to \infty} (-\{A,\Phi\})^N \ .
\end{equation}
We will also find that $P$ has a very close connection to the cohomology problem around the $\Phi$-vacuum and can be used to construct projector-like representatives of the cohomology of $\mathcal{Q}_\Phi$.

Given that all solutions are singular gauge transformations, it is tempting to try to find new solutions by constructing gauge transformations with certain singularities.  Following the model of the known cases, we suggest a new system for finding solutions and
give, as an example, a proposal for a lower-dimensional brane solution.  Since it is likely that there is still more to learn about the singularities of the gauge transformations in true solutions, our new solution is merely conjectural at this stage.  However, we hope that it will at least motivate further study in this direction.

The organization of the paper is as follows:  In section \ref{s:GaugeForm}, we demonstrate that every solution can be written in pure-gauge form and introduce to the notion of the characteristic projector to  explain why this does not trivialize all the known solutions. In section \ref{s:Agauge}, we discuss general aspects of cohomology and show how the characteristic projector can be used to generate projector-like representatives of the cohomology.  In section \ref{ChProjComp}, we apply our formalism to each of the known solutions.  In particular, we compute the characteristic projector for each solution. In section \ref{NewSolutions}, we speculate on how the formalism might be used to generate new solutions.  As an example, we propose a candidate lower-dimension brane solution.  In section \ref{s:OSFTequivalence}, we use the results of the previous sections to demonstrate that OSFT around the vacuum $\Phi$ is equivalent to OSFT defined around the BCFT associated with $\Phi$
\section{The pure-gauge form for general solutions} \label{s:GaugeForm}

In this section, we show how every solution around the tachyon vacuum can be written in pure-gauge form.
In the first subsection, we do this in a simple way using gauge transformations that do not satisfy the string-field reality condition.
Since the action is invariant under complex gauge transformations, this would suggest, if the gauge transformation were
non-singular, that the solution had the same energy as the tachyon vacuum.  In the next subsection, we show that there also exists
a manifestly real pure-gauge form of an arbitrary solution.

\subsection{The pure-gauge form of arbitrary solutions ignoring the reality condition}

Let $\Psi$ be the tachyon vacuum solution found in \cite{Schnabl:2005gv}.  The BRST operator
around the tachyon vacuum is given by
\begin{equation}
  \mathcal{Q} = Q_B + [\Psi,\cdot] \ .
\end{equation}
Let $\Phi$ be a solution of the equations of motion {\em around the tachyon vacuum},
\begin{equation}
  {\mathcal{Q}} \Phi + \Phi*\Phi = 0 \ .
\end{equation}
We would now like to show that $\Phi$ can always be written in pure-gauge form.
To do so, we recall the existence of the homotopy field $A$, which satisfies
\begin{equation}
  {\mathcal{Q}} A = 1 \ .
\end{equation}
The existence of $A$ implies that all {\em infinitesimal} solutions of the
equations of motion are trivial.  If ${\mathcal{Q}} \Phi = 0$, we have that $\Phi = {\mathcal{Q}}(A*\Phi)$.
Surprisingly, it is also enough to show that {\em all} solutions can be written in 
pure-gauge form\footnote{A formula which is similar to this one, but in a different context appeared in \cite{Fuchs:2008zx}.  This idea was also anticipated in \cite{Kwon:2008ap}, where marginal solutions of the tachyon vacuum were discussed.  The author has been informed that the formula presented here has been discovered in various forms by others \cite{BerkovitsSchnabl, ErlerSchnabl}}.   Consider that
\begin{equation}\label{QofOnePlusAPsi}
  {\mathcal{Q}}(1 + A* \Phi) = (1+ A*\Phi) * \Phi \ .
\end{equation}
Let $U = 1 + A*\Phi$, then (\ref{QofOnePlusAPsi}) implies
\begin{equation}
  {\mathcal{Q}} U = U\Phi \ ,
\end{equation}
so that
\begin{equation} \label{PG1}
  \Phi = U^{-1} {\mathcal{Q}} U \ .
\end{equation}
Note also that
\begin{equation}
  {\mathcal{Q}}(1+\Phi *  A) =- \Phi* \Phi A - \Phi = -\Phi*(1+\Phi*A) \ ,
\end{equation}
so that if $\widetilde{U} = 1+\Phi*A$, we have
\begin{equation}
  {\mathcal{Q}}\widetilde{U} = - \Phi* \widetilde{U},
\end{equation}
and
\begin{equation} \label{PG2}
  \Phi =-(\mathcal{Q} U) U^{-1} = \widetilde{U} {\mathcal{Q}} \widetilde{U}^{-1} \ .
\end{equation}
Assuming that $\Phi$ is not gauge-equivalent to the tachyon vacuum, (\ref{PG1}) and (\ref{PG2}) 
must in some way be anomalous. Since $U$ and $\widetilde{U}$ are finite products and sums of 
states we will assume that they are well-defined and that some problem must arise in the inverses $U^{-1}$ 
and $\widetilde{U}^{-1}$.   

The most obvious way that the inverse of an operator can be singular is if it has a non-trivial kernel.
If we found a left kernel for $U$, i.e. a $Z$ such $Z*U = 0$, this would imply that $Z *\Phi$ was divergent,
which is unpleasant, so we will assume that $U$ has a right kernel.  
Let us attempt, then, to solve the equation,
\begin{equation}
  U *Z =  0 \implies Z  = - A*\Phi * Z  \ .
\end{equation}
We learn that $A* Z = 0$. It follows that
\begin{equation}
   Z = (-\{A, \Phi\}) Z  \ .
\end{equation}
Defining $\Omega_\Phi = -\{A,\Phi\}$, we write this as
\begin{equation} \label{OmegaMultCondition}
  Z = \Omega_\Phi * Z \ .
\end{equation}
It follows that
\begin{equation}
  Z =(\Omega_\Phi)^N * Z
\end{equation}
for all $N$.
Now, suppose there is a limit,
\begin{equation} \label{PLim}
  P_\Phi = \lim_{N \to \infty} (\Omega_\Phi)^N \ .
\end{equation}
One possibility is $P_\Phi = 0$.  In this case, $Z = 0$ is the only solution and $U^{-1}$ is likely to be 
well-defined and the solution pure-gauge.  Given a physical solution $\Phi$ and taking $U = 1+ \lambda A*\Phi$,
with $\lambda \in [0,1]$, we see, then, that {\em every solution is formally a limit of a true gauge transformation of the tachyon vacuum}\footnote{Two comments on this statement are in order: First, for $\lambda$ less than $1$, the solution will not be real.  This problem will be fixed in the next subsection when we construct a manifestly real $U$.
Second, this statement would seem to rule out the possibility that $U$ is non-singular, but instead in some sense a ``large'' gauge transformation.  However, it may be that, in some string field norm, $1+\lambda A*\Phi$ is not close to $1+A*\Phi$ for $\lambda$ slightly smaller than $1$.  This happens, for instance in the $\mathcal{L}_0$ level truncation scheme \cite{Ellwood:2006ba}, although not in ordinary $L_0$ level truncation.  Note also that we cannot, at this time, rule out that $U^{-1}$ is a divergent operator even for $\lambda<1$, though in the known examples it is finite.}.

If $P_\Phi$ is non-vanishing and has  a finite product with itself, then, since it is a string field $\Omega_\Phi$ raised to an infinite power,
it will be a projector, which we call the characteristic projector.
\begin{equation}
  P_\Phi * P_\Phi = P_\Phi \ .
\end{equation}
Moreover, we learn that
\begin{equation} \label{ZProjCondition}
  P_\Phi*Z = Z \ ,
\end{equation}
which implies that the left half of $Z$ is invariant under the projection by left multiplication by $P_\Phi$.   The right half is left undetermined since the equation only involves left multiplication of $Z$ by $U$.

We can repeat the same discussion for $\widetilde{U}$ by searching for its  left kernel, $Z* \widetilde{U} = 0$.  The same assumptions about the limit in (\ref{PLim}) determine in this case that {\em right} half of $Z$ is invariant under the projection by {\em right} multiplication by $P_\Phi$.  

It will be convenient  later to group elements from the left kernel of $\widetilde{U}$ and right kernel of $U$ into a single string field $Z$ that satisfies (ignoring possible pathologies at the midpoint)
\begin{equation}
  U*Z =  Z* \widetilde{U} = 0 \ ,
\end{equation}
and, hence,
\begin{equation} \label{Zeq}
  Z = P_\Phi * Z = Z* P_\Phi \ .
\end{equation}
We learn that the left and right parts of $Z$ each live in a $k$-dimensional subspace where $k$ is the rank of $P_\Phi$ so that $Z$ lives in a $k^2$ dimensional subspace\footnote{This statement relies on an assumption that a projector will project the left and right half-string wave functions onto a subspace of dimension equal to the rank of the projector, which, in turn relies on  the non-degeneracy of the inner product between left and right half-string wave functions.  Careful consideration of the fine points of split string wave functions could make this claim considerably more subtle.  We will not dwell on this issue since, in all known examples, $P_\Phi$ is rank one.  We believe that higher rank characteristic projectors will only be important in multi-brane solutions, whose existence is currently rather conjectural.}.  The equations (\ref{Zeq}) determine the left and right parts of $Z$, but do not specify the behavior of $Z$ at the midpoint.  This allows one to introduce additional operators at the midpoint, which will play a role in discussion later.  In all the known examples, $P_\Phi$ is rank one so we find that the kernel is just given by $Z = P_\Phi$ which is the characteristic projector for $\Phi$ written in terms of either $U$ or $\widetilde{U}$.

Note that, in the case $Z = P_\Phi$, since $A*Z =Z*A = 0$, we have  $A*P_\Phi = P_\Phi*A = 0$.  The fact that $U$ has a right kernel and {\em not} a left kernel then arises from an associativity anomaly\footnote{The analogous anomaly for the tachyon vacuum was discovered in \cite{Ellwood:2006ba}.}.  Noting that
\begin{equation}
  P_\Phi *U = P_\Phi \ , \qquad \text{and} \qquad  U*P_\Phi = 0 \ ,
\end{equation}
we learn that
\begin{equation}
  (P_\Phi * U) * P_\Phi  = P_\Phi \ne  0  = P_\Phi * (U * P_\Phi) \ .
\end{equation}
giving a violation of associativity.

Before concluding this section, it is worth mentioning another possibility in the computation of $P_\Phi = \lim_{N \to \infty} (\Omega_\Phi)^N $: there is no limit.  This can happen, for instance, if we replace $U = 1 + A* \Phi \to 1 + \lambda A* \Phi$ with $\lambda >1$.  In this case $P_\Phi$ will diverge and it is unclear if there can be a kernel.  This situation arises, for example, in the pure-gauge form of the tachyon vacuum with an exotic gauge parameter.

In general, the series $P_N =  (\Omega_\Phi)^N $ can limit to a cycle or can diverge in a non-uniform way so that $(1-P_N) *Z$ remains small (or zero) for some $Z$, but diverges for others, still allowing for a kernel.  If the operator given by left multiplication by $\Omega_\Phi$ is diagonalizable, the existence of $P_\Phi$ relies on the assumption that all of its eigenvalues $\lambda$  satisfy  $|\lambda| \le 1$ with the additional requirement that if $|\lambda| = 1$, then $\lambda = 1$. The existence of $P_\Phi$ for all the known solutions suggests that this may be a general rule.  We suspect that all of the eigenvalues should be real since $\Omega_\Phi$ is a real string field, $\Omega_\Phi^\ddagger = \Omega_\Phi$, however we do not have a sound argument for this claim.  Additionally, we suspect that eigenvalues $|\lambda|>1$ would lead to operators $U^{-1}$ which are too singular.  If, in general, one needs to consider $\Omega_\Phi$ with eigenvalues $|\lambda|>1$, one can still define $P_\Phi$ as the projector onto the space of fields with eigenvalue one; however, it would be considerably more difficult to find $P_\Phi$ in this case.

From the perspective of level truncation, if we have $\Phi = U^{-1} * QU$, we must balance two competing concerns.  On the one hand, $U^{-1}$ must not be a true inverse of $U$.  This suggests that $U^{-1}$ should not fall off too quickly at high levels.  On the other hand, the solution $\Phi$ must be finite, so $U^{-1}$ should not grow too quickly at high levels.   If we found that $U$ had a kernel in the Fock-space, we would expect that $U^{-1}$ would diverge in the Fock-space expansion.  If, instead, $U$ only had a kernel consisting of fields which grew rapidly at high level, this might not be enough to rule out the existence of a well-defined inverse, $U^{-1}$.    Speaking loosely, we expect that, for a proper solution of the equations of motion, the kernel of $U$ should consist of fields which neither fall off too quickly at high level, nor grow too quickly.
Projectors, being just outside the space of wedge states with insertions, are natural candidates for elements of the kernel of $U$.

\subsection{A manifestly real pure-gauge form of general solutions}

The string-field reality condition $\Phi^{\ddagger} = \Phi$ \cite{Gaberdiel:1997ia} implies that a real gauge transformation should satisfy
\begin{equation}
  U^{\ddagger} = U^{-1} \ .
\end{equation}
Noting that $A^{\ddagger} = A$ and $1^{\ddagger} = 1$, we see that
\begin{equation}
  (1+A*\Phi)^{\ddagger} = (1+\Phi*A) \ ,
\end{equation}
however, $(1+\Phi*A)^{-1} \ne (1+A*\Phi)$.  This implies that the gauge transformations of the previous section are not real (though $U$ and $\widetilde{U}$ are conjugates of each other under $\ddagger$).

To find a real gauge transformation, it is convenient to work with
\begin{equation}
  \Omega^L =- A* \Phi \ , \qquad \Omega^R = -\Phi*A \ .
\end{equation}
Note that $\Omega^R \Omega^L = 0$, so that, in any power-series expansion in $\Omega$s, all the $\Omega^L$ will be to the left of all the $\Omega^R$.
We search for a gauge transformation of the form,
\begin{equation}
  U =  \nol f(\Omega^L,\Omega^R) \nor \ .
\end{equation}
The $\nol$ are used to denote  a normal ordering in which we move all the $\Omega^L$s to the left of the $\Omega^R$s
before evaluating the expression.  For example, we have
\begin{equation}
  \nol \Omega^R \Omega^L \nor = \Omega^L \Omega^R \ .
\end{equation}
As a more complicated example, we put
\begin{equation}
  \nol \Omega^R\Omega^L \Omega^R + \Omega^L \Omega^R\Omega^L \nor = 
  \Omega^L (\Omega^R)^2 + (\Omega^L)^2 \Omega^R \ .
\end{equation}
In general, an expression $\nol h(\Omega^L, \Omega^R) \nor$ will not be independent of $\Omega^{L,R}$ unless $h(x,y)$ is a constant.

Note that with this prescription,
\begin{equation}
  (\nol f(\Omega^L ,\Omega^R) \nor )^{\ddagger} = \nol f(\Omega^R ,\Omega^L) \nor  \ .
\end{equation}
It follows that the reality condition is given by
\begin{equation} \label{Unitarity}
  \nol f(\Omega^R, \Omega^L)\nor * \nol f(\Omega^L \Omega^R)\nor = 1 \ .
\end{equation}
To show that $\Phi$ is formally pure-gauge, we wish to find such an $f$ that also satisfies
\begin{equation}\label{Puregaugeitude}
  \nol f(\Omega^R, \Omega^L)\nor *\mathcal{Q}( \nol f(\Omega^L, \Omega^R)\nor)  = \Phi \ .
\end{equation}
To make progress, we reduce (\ref{Unitarity}) and (\ref{Puregaugeitude}) to ordinary equations in commutative algebra for $f$.
First, we start with  (\ref{Unitarity}).  Noting again that $\Omega^R * \Omega^L = 0$, if we multiply two functions of 
$\Omega^{L,R}$, we get a non-zero answer only when either the left factor has no $\Omega^R$s or the right factor has no factors of
$\Omega^L$s.  Hence (\ref{Unitarity})   reduces to
\begin{equation}
  \nol  f(0,\Omega^L) f(\Omega^L,\Omega^R) + f(\Omega^R,\Omega^L) f(0,\Omega^R) - f(0,\Omega^L) f(0,\Omega^R)\nor  = 1 \ .
\end{equation}
It follows that,
\begin{equation} \label{fEq1}
  f(0,x) f(x,y) + f(y,x)f(0,y) - f(0,x) f(0,y) = 1 \ .
\end{equation}

To study the second identity (\ref{Puregaugeitude}) we make use of the following relations:
\begin{equation}
  \mathcal{Q} \Omega^L = -(1-\Omega^L)*\Phi \ , \qquad \mathcal{Q} \Omega^R =  \Phi* (1-\Omega^R) \ .
\end{equation}
We also have
\begin{equation}  
  \Phi *\Omega^L = \Omega^R*\Phi \ .
\end{equation}
To define a useful normal ordering scheme for products of $\Omega^{L,R}$s and one $\Phi$, we need to introduce extra symbols
which specify whether each $\Omega^{L,R}$ is to the left or right of $\Phi$.  We do this with the subscripts $\ell,r$.  For example,
we put
\begin{equation}
   \nol  \Phi \Omega^{R}_\ell \Omega^L_\ell  \Omega^L_r \nor  = \Omega^L \Omega^R \Phi \Omega^L \ .
\end{equation}
Consider, then,
\begin{multline}
  \mathcal{Q} \nol  (\Omega^L)^N (\Omega^R)^M \nor  = \mathcal{Q} ((\Omega^L)^N (\Omega^R)^M)
  \\
   = -\sum_{k = 0}^{N-1} (1-\Omega^L)  (\Omega^L)^k \Phi (\Omega^L)^{N-k-1} (\Omega^R)^M 
    + \sum_{k=0}^{M-1}  (\Omega^L)^N (\Omega^R)^{k} \Phi (\Omega^R)^{M-k-1} (1-\Omega^R) \ ,
\end{multline}
which we rewrite as
\begin{equation}
-\sum_{k = 0}^{N-1} (1-\Omega^L)  (\Omega^L)^k (\Omega^R)^{N-k-1}\Phi (\Omega^R)^M 
    + \sum_{k=0}^{M-1}  (\Omega^L)^N\Phi (\Omega^L)^{k} (\Omega^R)^{M-k-1} (1-\Omega^R) \ .
\end{equation}
Using our normal ordering scheme, this is equivalent to
\begin{equation}
  -\sum_{k = 0}^{N-1} \nol  (1-\Omega^L_\ell)  (\Omega^L_\ell)^k (\Omega^R_\ell)^{N-k-1}\Phi (\Omega^R_r)^M  \nor 
    + \sum_{k=0}^{M-1}  \nol  (\Omega^L_\ell)^N\Phi (\Omega^L_r)^{k} (\Omega^R_r)^{M-k-1} (1-\Omega^R_r)\nor  \ .
\end{equation}
We can now perform the sum over $k$ to get
\begin{equation}
  \nol (\Omega^L_\ell-1) \frac{(\Omega^L_\ell)^{N} -(\Omega^R_\ell)^{N}}{\Omega^L_\ell - \Omega^R_\ell} \Phi(\Omega_r^R)^M
   + (\Omega^L_\ell)^N \Phi \frac{(\Omega^L_r)^{M} -(\Omega^R_r)^{M}}{\Omega^L_r - \Omega^R_r} (1-\Omega^R_r)\nor  \ .
\end{equation}
Assuming that $f(\Omega^L,\Omega^R)$ has a Taylor series expansion in powers of $\Omega$s, we learn that
\begin{multline}
  \mathcal{Q} \nol  f(\Omega^L,\Omega^R) \nor  \\
   = \nol (\Omega^L_\ell-1) \frac{f(\Omega^L_\ell,\Omega^R_r) - f(\Omega^R_\ell,\Omega^R_r)}{\Omega^L_\ell - \Omega^R_\ell} \Phi
   + 
  \Phi \frac{f(\Omega^L_\ell,\Omega^L_r) -f(\Omega^L_\ell,\Omega^R_r)}{\Omega^L_r - \Omega^R_r} (1-\Omega^R_r)\nor  \ .
\end{multline}
We can push the $\Phi$ as far to the right as possible by replacing $\Omega^L_r \to \Omega^R_\ell$.  This gives
\begin{multline} \label{QOnGauge}
  \mathcal{Q} \nol  f(\Omega^L,\Omega^R) \nor  \\
   = \nol (\Omega^L_\ell-1) \frac{f(\Omega^L_\ell,\Omega^R_r) - f(\Omega^R_\ell,\Omega^R_r)}{\Omega^L_\ell - \Omega^R_\ell} \Phi
   + 
  \Phi \frac{f(\Omega^L_\ell,\Omega^R_\ell) -f(\Omega^L_\ell,\Omega^R_r)}{\Omega^R_\ell - \Omega^R_r} (1-\Omega^R_r)\nor  \ .
\end{multline}
To complete the computation of (\ref{Puregaugeitude}), we multiply on the left by $\nol f(\Omega^R_\ell,\Omega^L_\ell)\nor $.  Let the right hand side of (\ref{QOnGauge}) be given by $\nol g(\Omega^L_\ell,\Omega^R_\ell,\Omega^R_r) \Phi\nor $ so that
\begin{equation} \label{fEq2}
  g(x,y,z) = (x-1) \frac{f(x,z)-f(y,z)}{x-y} + (1-z) \frac{f(x,y) - f(x,z)}{y-z} \ .
\end{equation}  
Then  (\ref{Puregaugeitude}) is equivalent to
\begin{equation}\label{fEq3}
  f(0,x) g(x,y,z) + f(y,x) g(0,y,z) -f(0,x)g(0,y,z) = 1 \ .
\end{equation}
We have not found a way to algebraically solve (\ref{fEq1}, \ref{fEq2}, \ref{fEq3}) for $f(x,y)$. However, we were able to guess the following  analytic expression:
\begin{equation} \label{fForm}
  f(x,y) = \frac{1-x}{x-y} \left(\frac{x}{\sqrt{1-x}} - \frac{y}{\sqrt{1-y}} \right) \ ,
\end{equation}
which the reader may verify satisfies (\ref{fEq1}, \ref{fEq2}, \ref{fEq3})\footnote{Upon seeing a draft of this paper, another form of this gauge transformation was suggested by Erler.  This is presented in appendix \ref{PGF} along with a proof and some additional identities.} .

An explicit Taylor-series expansion is given by
\begin{equation} \label{Uformula}
  U = \nol f(\Omega^L,\Omega^R)\nor  = \sum_{m,n = 0}^\infty a_{m,n} (\Omega^L)^m(\Omega^R)^n \ ,
\end{equation}
where
\begin{equation}
   a_{m,0} = - \frac{\Gamma(m-1/2)}{2\sqrt{\pi} \,\Gamma(m+1)} \ ,
\end{equation}
together with $a_{m,n} = a_{m+n, 0}$ for $m \ne 0$ and $a(0,n) = (1-2n) a(n,0)$.

Due to the complexity of the real gauge transformation, we have not been able to give as complete an account of
when it becomes singular.  However, the following facts suggest that the story is similar.  Consider $Z$ to be annihilated
by $A$.  Then we have
\begin{equation}
  \nol  f(\Omega^L,\Omega^R) \nor  *Z = \nol  f(\Omega^L,0)\nor  *Z = \sqrt{1-\Omega^L} * Z \ .
\end{equation}
Hence, we find that the same characteristic projector found before is a right kernel of the gauge transformation.

Similarly, if $Z$ is annihilated from the right by $A$, we have
\begin{equation}
  Z* \nol f(\Omega^L,\Omega^R)\nor  = Z* \frac{1}{\sqrt{1-\Omega^R}} \ .
\end{equation}
It follows that the gauge transformation diverges when multiplied from the left by the characteristic projector.  Note that,
unlike in the previous section, the gauge transformation and its inverse are both ill-defined for non-trivial solutions.  This 
is required by the reality condition.

\section{Cohomology and the characteristic projector} \label{s:Agauge}

In this section, we discuss the close relationship between the characteristic projector and the cohomology
of the BRST operator around the shifted vacuum.  As in the previous section,
$\mathcal{Q}$ will denote the BRST operator around the tachyon vacuum and  $A$ will denote the homotopy field,
\begin{equation} \label{QA}
  \mathcal{Q} A = 1 \ .
\end{equation}

Using $A$, we can form two operators, $\ell_A$ and $r_A$, which are
left and right multiplication by $A$,
\begin{equation}
  \ell_A \Sigma = A* \Sigma \ , \qquad r_A \Sigma = (-1)^{\Sigma A} \Sigma * A \ .
\end{equation}
These operators obey 
\begin{equation}\label{rlidentities}
  \{\mathcal{Q} , \ell_A \} = \{\mathcal{Q} , r_A \} = 1 \ , \qquad \{ \ell_A, r_A \} = 0 \ , 
  \qquad\text{and} \qquad  r_A^2 = \ell_A^2 = 0 \ .
\end{equation}
As is standard in cohomology discussions, if we anticommute
a generic BRST operator $Q$ with a ghostnumber $-1$ operator, $\mathcal{O}$ the
resulting ghostnumber $0$ operator, $K = \{Q,\mathcal{O}\}$,
if it is diagonalizable, contains a representative of entire cohomology of
$Q$ in its kernel.  To see this, suppose $Q\Sigma = 0$
and $K\Sigma = \lambda \Sigma$, with $\lambda \ne 0$.  Then
\begin{equation}
  Q (\lambda^{-1} \mathcal{O} \Sigma) = \lambda^{-1} K \Sigma = \Sigma \ ,
\end{equation}
so that $\Sigma$ is exact.
Examining, (\ref{rlidentities}), we see that, since $1$ has no kernel,
the cohomology of $\mathcal{Q}$ must vanish.

Given $\Phi$ solving $\mathcal{Q} \Phi + \Phi*
\Phi = 0$, the BRST operator in the $\Phi$-vacuum
is given by $\mathcal{Q}_{\Phi} \Sigma = \mathcal{Q}\Sigma +
\Phi * \Sigma - (-1)^\Sigma \Sigma *\Phi$.  Consider the
analogue of (\ref{QA}) in this vacuum,
\begin{equation} \label{eq:QonA}
  \mathcal{Q}_{\Phi} A = 1 - \Omega_{\Phi} \ ,
\end{equation}
where $\Omega_{\Phi} = - \{\Phi, A\}$.  Acting on both
sides of (\ref{eq:QonA}) with $\mathcal{Q}_{\Phi}$, we learn
that
\begin{equation} \label{QOmega}
  \mathcal{Q}_{\Phi} \Omega_{\Phi} = 0 \ .
\end{equation}
Note that if $\Omega_{\Phi} = \mathcal{Q}_{\Phi} \Lambda$,
we would find $\mathcal{Q}_{\Phi}(A+\Lambda) = 1$ and the
cohomology of $\mathcal{Q}_{\Phi}$ would be trivial.  It follows
that, if $\Phi$ is not pure-gauge (i.e. is {\em not} the tachyon
vacuum), then $\Omega_{\Phi}$ is in the cohomology of
$\mathcal{Q}_{\Phi}$.

From (\ref{eq:QonA}), we learn that
\begin{equation}
  \{\mathcal{Q}_{\Phi}, r_A\} = 
   1 - r_{\Omega_{\Phi}}  = r_{\widetilde{U}}\ , 
\qquad 
  \{\mathcal{Q}_{\Phi}, \ell_A\} = 1 - \ell_{\Omega_{\Phi}}  = \ell_U\ .
\end{equation}
Assuming that $r_{\widetilde{U}}$ and
$\ell_{U}$ are diagonalizable operators, and noting
that they commute, we can assume that an entire representative of the cohomology of
$\mathcal{Q}_{\Phi}$ lives in the intersection of their kernels:
$\text{coh}(\mathcal{Q}_{\Phi}) \subset
\text{ker}(r_{\widetilde{U}}) \cap
\text{ker}(\ell_U)$.  Consider a ghostnumber
zero state $Z$ which lives in this space.  

We see that such a $Z$ will obey the same condition we found earlier
\begin{equation}\label{eq:Zrel}
U*Z = Z*\widetilde{U} = 0 \ .
\end{equation}
With the same assumptions that we used in the previous section, the solutions to this equation
are given by the condition that the characteristic projector $P_\Phi$ leaves $Z$ invariant
when multiplying from the left or right.  Hence, modulo midpoint insertions, $Z$ lives in 
$k^2$-dimensional subspace where $k$ is the rank of $P_\Phi$.

In the known examples that we will study shortly, $k = 1$ and we see that $Z = P_\Phi$
up to midpoint insertions.  At ghostnumber zero, 
we will generally assume that there are no such operators except the identity, as is usually the
case in critical string theory, so that $Z = P_{\Phi}$ is the
unique element of the ghost number zero cohomology.  There could
be exceptions to this rule in non-critical strings in low
dimensions.  We thus find that the characteristic projector is typically an element of 
the ghost number zero cohomology.  This is not surprising considering that it is given
by $P_\Phi = \lim_{N\to \infty} \Omega_\Phi^N$, where $\Omega_\Phi$ was shown earlier
to be an element of the cohomology.

We can also construct higher ghostnumber elements of the cohomology.  These are found
by inserting weight $(0,0)$ primaries at the midpoint of $Z$ which does not affect the conditions 
$U*Z = Z*\widetilde{U} = 0$.  Conveniently, we will find in the examples that the boundary conditions
near the midpoint of the projector $Z$ are precisely those of the boundary CFT dual to the solution $\Phi$.
Since a complete collection of representatives of the cohomology can be found among the weight $(0,0)$
primaries, we can cover the entire expected cohomology in this way.

We should emphasize that, in spite of the simplicity of this construction, it is still desirable to construct non-projector-like representatives
of the cohomology.  We have only been able to do so in the case of marginal deformations with trivial OPE (see appendix \ref{s:NonProjCohomology}) using completely different methods.  Nonetheless, we find the close connection between 
the characteristic projector and cohomology to be an important hint of their role in the structure of SFT solutions. 

It is very suggestive that, when the rank of $P_\Phi$ is $k>1$, we can find $k^2$ projectors $Z$ in which 
to insert operators at the midpoint.  These many copies of the cohomology hint that higher rank characteristic
projectors correspond to multiple brane solutions with $k$ the number of branes. Since no multiple-brane solutions have been constructed we will not pursue this
idea, but we nonetheless find it intriguing.

\section{The characteristic projector of the known solutions} \label{ChProjComp}

In this section we compute the characteristic projector for each of the known solutions.  The existence of
this projector is essential to ensuring that these solutions are not gauge transformations of the tachyon vacuum.

\subsection{The perturbative vacuum}

In this section we study the simplest example we can apply this formalism to: the case
$\Phi = - \Psi$, or the perturbative vacuum.  Note that since
we are writing everything from the perspective of the tachyon vacuum
state, the perturbative vacuum is a non-trivial solution.  Using the
identity found in \cite{Ellwood:2006ba},
\begin{equation}
  \Psi * A + A * \Psi = |0\rangle \ ,
\end{equation}
we learn that
\begin{equation}
  \Omega_{\Phi} = -\{\Phi,A\} = |0\rangle \ .
\end{equation}
so that the characteristic projector is given by
\begin{equation}
  P_{\Phi} = \lim_{M\to \infty} \Omega_{\Phi}^M = \lim_{M\to \infty} |0\rangle ^M = |\infty \rangle.
\end{equation}
As expected, this state is an element of the cohomology of
$\mathcal{Q}_{\Phi} = Q_B$.  We can construct higher ghost
number elements in the kernel of $\ell_U$ and $r_{\widetilde{U}}$ by
inserting weight $(0,0)$ primaries at the midpoint, which, since we
are working with a projector, is also the boundary of the worldsheet.
Such operators nicely fill out the rest of the cohomology of $Q_B$.

Note that we have not associated the tachyon vacuum state with sliver,
but the {\em perturbative vacuum}.  In the notation we are using in
this section, the tachyon vacuum is given by the state ${\Phi} =
0$ and hence $\Omega_{\Phi} = P_{\Phi} = 0$.

\subsection{Marginal deformations with trivial OPE} \label{s:trivialOPE}

For marginal deformations with trivial OPE, a $\mathcal{B}_0$-gauge
solution was found in \cite{Schnabl:2007az,Kiermaier:2007ba}.  It is
given by
\begin{equation}
  \Theta = -\Psi + \Theta_0 = -\Psi -W_{1/2} * \frac{1}{1 - \lambda  \, cJ  * A} * \lambda\, cJ  * W_{1/2} \ .
\end{equation}
Here we are using the following notation: $W_r$ is the wedge state \cite{Rastelli:2000iu} of
width $r$ (with $r = 1$ the $SL_2(\mathbb{R})$ vacuum) .  As before,
$A$ is the homotopy field, which is used here for convenience.  The
operator $J$ denotes a weight $1$ boundary primary and $c$ is the
$c$-ghost.  As a string field, $cJ$ is taken to be the operator $cJ$
inserted at the boundary of the identity string field, $W_0$.

The state $\Omega_{\Theta}$ is given by
\begin{equation}
\Omega_{\Theta} =   |0\rangle - \{ A,\Theta_0\} \ .
\end{equation}
With this state in hand, we can search for the projector,
\begin{equation}\label{PProd}
  P_{\Theta} = \lim_{M \to \infty} \Omega_{\Theta}^M = 
   \lim_{M\to \infty} \left(|0\rangle - \{A,\Theta_0\} \right)^M \ .
\end{equation}
Although the computation of $P_{\Theta}$ is somewhat involved,
the final result given in (\ref{MarginalPFinalAnswer}) is very simple and readers
interested only in the results can feel free to skip ahead.

As $M$ becomes large in the product (\ref{PProd}) the width of the
worldsheets that make up $\Omega^M$ will become large and, to find the
limit, we should either focus on the left end of the worldsheet or the
right end.  From an algebraic point of view, this implies that we
should focus only either the left or right end of the long string of fields being
multiplied together.

For example, keeping only terms up to $\mathcal{O}(\lambda^1)$,
$P_{\Theta}$ takes the form (as viewed from the left end of the
string of multiplications or, equivalently, the left side of the
string worldsheet),
\begin{equation}
  P_{\Theta}\overset{\text{left-half}}{=} |\infty \rangle 
  + \lambda \sum_{N = 0}^\infty |0\rangle^N *A * W_{1/2} * c J * |\infty\rangle
  + \mathcal{O}(\lambda^2) \ .
\end{equation}
We will use the notation $\overset{\text{left-half}}{=}$ and $\overset{\text{right-half}}{=}$
to refer to equalities that only hold for the left and right halves of the string wave function.
Now, using the fact that
\begin{equation}
  \sum_{N = 0}^\infty |0\rangle^N *A = \int_0^\infty dr\, W_r B_1^L \ ,
\end{equation}
we find
\begin{equation}
 P_{\Theta} \overset{\text{left-half}}{=} 
 |\infty \rangle + \lambda \int_0^\infty dr\, W_{r+1/2} *J * |\infty\rangle
 +\mathcal{O}(\lambda^2) \ .
\end{equation}
Note that this expression is only correct for the {\em left half of
the state}, since we are ignoring the right hand side of the chain of
operators in the infinite product.  Focusing instead on the right half, one would find
\begin{equation}
 P_{\Theta} \overset{\text{right-half}}{=} 
 |\infty \rangle +  |\infty\rangle
* J* \lambda \int_0^\infty dr\, W_{r+1/2}
 +\mathcal{O}(\lambda^2) \ .
\end{equation}
It follows that the first order expression for $P_\Theta$ is given by
\begin{equation}
  P_\Theta = |\infty \rangle + \lambda \int_0^\infty dr\, W_{r+1/2} *J * |\infty\rangle
 + |\infty\rangle
* J* \lambda \int_0^\infty dr\, W_{r+1/2}
 +\mathcal{O}(\lambda^2) \ .
\end{equation}

The complete computation to all orders in $\lambda$ gives (viewing
everything from the left),
\begin{align}
P_{\Theta} \overset{\text{left-half}}{=}
  \sum_{N = 0}^\infty \left(\int_0^\infty dr\, W_{r+1/2} \sum_{K = 0}^\infty \left( 
    \lambda\, J \int_0^1 dr' W_{r'}
  \right)^K *\lambda\, J * W_{1/2}\right)^N *|\infty\rangle \ .
\end{align}
To keep the formulae compact, let
\begin{equation}
  W_a^b = \int_a^b dr\, W_r \ ,
\end{equation}
so that
\begin{align}\label{fullsum}
P_{\Theta} \overset{\text{left-half}}{=}
  \sum_{N = 0}^\infty \left(W_{1/2}^\infty \sum_{K = 0}^\infty \left( 
    \lambda\, J* W_0^1
  \right)^K *\lambda\, J * W_{1/2}\right)^N *|\infty\rangle \ .
\end{align}
Noting that
\begin{equation}
  (A*B)^N = A*(B*A)^{N-1}* B \ ,
\end{equation}
we can rewrite (\ref{fullsum}) as
\begin{align}
  |\infty\rangle +  W_{1/2}^\infty \frac{1}{1- \lambda J *W_0^1} \sum_{N = 0}^\infty
  \left(
    \lambda J * W_1^\infty \frac{1}{1- \lambda J *W_0^1} 
  \right)^N
  \lambda J * |\infty\rangle \ ,
\end{align}
which simplifies to
\begin{multline}
  |\infty\rangle + W_{1/2}^\infty \frac{1}{1- \lambda J *W_0^1}   \left[\frac{1}{
   1- \lambda J * W_1^\infty (1- \lambda J *W_0^1)^{-1}
  } \right]
  \lambda J * |\infty\rangle
  \\
  =
  |\infty \rangle
   +  W_{1/2}^\infty \frac{1}{1- \lambda J *W_0^\infty} \lambda J*|\infty\rangle 
   \\
   =
   |\infty \rangle + W_{1/2} *\frac{\lambda W_0^\infty *J}{1- \lambda W_0^\infty *J} *|\infty\rangle
    = W_{1/2} * \frac{1}{1- \lambda W_0^\infty *J} * |\infty \rangle \ .
\end{multline}
Expanding this out, our result is 
\begin{equation}\label{eq:PR}
  P_\Theta \overset{\text{left-half}}{=} W_{1/2} *\sum_{N = 0}^\infty
  \left(\int_0^\infty dr\,W_r* \lambda J \right)^N *|\infty\rangle \ .
\end{equation}
Viewed from the right, we would have found
\begin{equation}\label{eq:PL}
  P_{\Theta} \overset{\text{right-half}}{=}
  |\infty\rangle * \sum_{N = 0}^\infty \left(\int_0^\infty dr\,J*W_r\right)^N * W_{1/2} \ .
\end{equation}
Together, (\ref{eq:PR}) and (\ref{eq:PL}) determine the right and left
halves of the projector $P_{\Theta}$.  We see that it is given
by
\begin{equation}\label{TrivialOPEP}
 P_\Theta = W_{1/2} *\sum_{N = 0}^\infty
\left(\int_0^\infty dr\,W_r* \lambda J \right)^N *|\infty\rangle * \sum_{N = 0}^\infty \left(\int_0^\infty dr\,J*W_r\right)^N * W_{1/2} \ .
\end{equation}
In CFT language, this is a very simple object:
\begin{equation}  \label{MarginalPFinalAnswer}
  \langle \phi | P_{\Theta} \rangle = 
  \left \langle f\circ \phi  \exp\left( \int_{1}^\infty dy\, \lambda J(y) \right) \exp\left(\int_{-\infty}^{-1} dy, \lambda J(y) \right) \right\rangle_{\text{UHP}} \ ,
\end{equation}
where $f = \frac{2}{\pi} \arctan(z)$.  This reorganization of the
complicated integrals over the positions of the $J$'s in $\Theta$ into
a simple exponential in $P_\Theta$ is quite remarkable and is
reminiscent of a similar phenomenon in \cite{Ellwood:2008jh,Kiermaier:2008qu}.  As we
will see in non-trivial OPE case, this is by no means a requirement.
This state appeared in \cite{Kiermaier:2007vu} and, in their notation,
we have the main result of this section:
\begin{equation} \label{PPhi}
  \vphantom{\Bigl(} \hphantom{\text{l}} P_\Theta = U_\infty \ .
\end{equation}

Another simple way of writing this state down is to let $|
\infty_{\lambda J} \rangle$ be the sliver state, but with boundary
conditions given by BCFT${}_{\Theta}$, the boundary conformal field
theory associated with marginal deformation $\Theta$.  Then
$P_{\Theta} = W_{1/2} * |\infty_{\Theta} \rangle * W_{1/2}$.  In
other words, except for two strips of worldsheet at each end of
$P_\Theta$, the projector is just the sliver, but with new boundary
conditions.  Note that, as expected, $P_\Theta * P_\Theta =
P_\Theta$.  That this product is finite requires that $J$ has a
trivial OPE, since, otherwise, the operator collisions would produce a
divergence.

\subsubsection{Cohomology on $P_{\Theta}$}
Here we discuss how $P_\Theta$ can be used to make elements of the BRST cohomology.  It is also possible to construct 
non-projector-like representatives, and the interested reader should refer to appendix \ref{s:NonProjCohomology} for details.

It is instructive to check first that, as expected from (\ref{QOmega}), $\mathcal{Q}_{\Theta} P_{\Theta} = 0$, which reduces to
\begin{equation}
  Q_B P_{\Theta} + \{ \Theta_0 ,P_{\Theta}\} = 0 \ .
\end{equation}
We have
\begin{equation} \label{QPPhi}
  \langle \phi| Q_B P_{\Theta}\rangle =
  \left \langle f\circ \phi \left( cJ(-1) - cJ(1) \right)
   \exp\left( \int_{1}^\infty dy\, \lambda J(y) \right) \exp\left(\int_{-\infty}^{-1} dy, \lambda J(y) \right) \right\rangle_{\text{UHP}} \ .
\end{equation}
As expected, since $P_{\Theta}$ is a projector, $Q_B
P_{\Theta}$ can be thought of as $Q_B$ acting separately on the
left half of $P_{\Theta}$, giving the $cJ(-1)$ term and on the
right half, giving the $-cJ(1)$ term.  These two pieces cancel with
$\Theta * P_{\Theta}$ and $P_{\Theta}* \Theta$
respectively.

Consider (keeping track of only the left half of $P_{\Theta}$),
\begin{equation}
  \Theta_0 * P_{\Theta} = -W_{1/2} * c J * \frac{1}{1-W_0^1 * J} *W_1 * \sum_{N = 0}^\infty \left(W_0^\infty *J \right)^N *|\infty\rangle \ ,
\end{equation}
which we can rewrite as
\begin{equation} \label{ThetaP}
  -W_{1/2} * cJ * \sum_{K = 0}^\infty \sum_{L = 0}^K \left(W_0^1 *J\right)^L *W_1*
  \left( W_0^\infty * J\right)^{K-L} * |\infty\rangle \ .
\end{equation}
It is straightforward to check by induction on $K$ that
\begin{equation}\label{sumidentity}
  \sum_{L = 0}^K \left(W_0^1 *J\right)^L *W_1*
  \left( W_0^\infty * J\right)^{K-L} *|\infty \rangle
   = \left( W_0^\infty * J\right)^{K} *|\infty\rangle \ .
\end{equation}
For $K = 1$, (\ref{sumidentity}) reduces to the trivial identity,
\begin{equation}
  \left(W_0^1 *J\right) *|\infty\rangle + W_1*\left( W_0^\infty * J\right)*|\infty\rangle
   = W_0^\infty *J * |\infty\rangle \ .
\end{equation}
Assuming that the identity holds for $K-1$, (\ref{sumidentity}) reduces to
\begin{equation}
  \left(W_0^1 *J\right)*\left( W_0^\infty *J\right)^{K-1} *|\infty\rangle
  + W_1 * \left( W_0^\infty *J\right)^{K} *|\infty\rangle
  =
    \left( W_0^\infty * J\right)^{K} *|\infty\rangle \ ,
\end{equation}
which, again holds trivially.  This proves (\ref{sumidentity}).  Equation (\ref{ThetaP}) reduces to
\begin{equation}
  \Theta_0*P_\Phi =  -W_{1/2} * cJ * \sum_{K=0}^\infty
  \left( W_0^\infty * J\right)^{K} * |\infty\rangle \ ,
\end{equation}
or, equivalently,
\begin{equation}
\langle \phi |\Theta_0*P_\Phi \rangle = 
-\left \langle f\circ \phi\,\,   cJ(-1)  
   \exp\left( \int_{1}^\infty dy\, \lambda J(y) \right)
 \exp\left(\int_{-\infty}^{-1} dy, \lambda J(y) \right) \right\rangle_{\text{UHP}} \ ,
\end{equation}
which cancels the $cJ(-1)$ term in (\ref{QPPhi}).  The other term is
cancelled by $P_\Phi*\Theta$.  It follows that
\begin{equation} \label{QPPhi2}
  \mathcal{Q}_\Theta P_\Phi = Q_B P_\Phi + \Theta_0 * P_\Phi + P_\Phi *\Theta_0 = 0 \ .
\end{equation}
Now, consider putting a weight $(0,0)$ primary $\mathcal{V}$ at the
midpoint of $P_\Phi$ to form a new state $P_\Phi(\mathcal{V})$,
\begin{equation}
 \langle \phi|  P_\Phi (\mathcal{V}) \rangle = 
 \left \langle f\circ \phi \,\,  \mathcal{V}(i \infty)
 \exp\left( \int_{1}^\infty dy\, \lambda J(y) \right) \exp\left(\int_{-\infty}^{-1} dy, \lambda J(y) \right) \right \rangle \ .
\end{equation}
Note that $\mathcal{V}$ is being inserted on the boundary of the disk
(at $i\infty$ in UHP coordinates) and, hence, must be a weight $(0,0)$
primary in BCFT${}_J$.  We also have, using (\ref{QPPhi2}),
\begin{equation}
  \mathcal{Q}_\Phi P_\Phi(\mathcal{V}) = P_\Phi(Q_B \mathcal{V}) \ .
\end{equation}
It follows that the cohomology computation on the restricted set of
states $P_\Phi(\mathcal{V})$ reduces to the standard cohomology of the
BRST operator on weight $(0,0)$ primaries in BCFT${}_J$.  Such
operators, conveniently, contain the entire ghostnumber one cohomology of the BRST
operator in terms of states of the form $c \mathcal{O}$ where
$\mathcal{O}$ is a weight $(1,1)$ primary.

\subsection{Marginal deformations with non-trivial OPE} \label{s:nontrivialOPE}

A solution representing marginal transformations with non-trivial OPE,
\begin{equation}
  J(z) J(0) \sim \frac{1}{z} \ ,
\end{equation}
was constructed in \cite{Fuchs:2007gw,Fuchs:2007yy,Kiermaier:2007vu}.
Here we use the form developed in \cite{Kiermaier:2007vu}. In spite of the apparent
complexity of the solution, the computation is actually easier than
in the trivial OPE case.

\subsubsection{Review of the Kiermaier-Okawa form of the non-trivial OPE marginal solution}

The solution is constructed from solutions $\Upsilon_{L,R}$ which do
not satisfy the OSFT reality condition,
\begin{equation}
  \Upsilon_{L} =-\Psi+\Upsilon_{0L} = -\Psi  -A_L U^{-1} \ , \qquad \Upsilon_R = -\Psi +\Upsilon_{0R} =  -\Psi -U^{-1} A_R \ ,
\end{equation}
where the fields $A_{L,R}$ and $U$ are defined in
\cite{Kiermaier:2007vu}\footnote{Unfortunately,
\cite{Kiermaier:2007vu} uses the opposite convention for the left and
right half of the string wave function, so we have
$A_{L,R}^{\text{here}} = A_{R,L}^{\text{there}}$}.  Their definitions
will be repeated shortly for convenience. The complete solution is
given by 
\begin{equation}
  \Upsilon = -\Psi +\Upsilon_0 = \frac{1}{\sqrt{U}} \left( \Upsilon_L + Q_B\right) \sqrt{U} = \sqrt{U} \left(\Upsilon_R +Q_B \right) \frac{1}{\sqrt{U}} \ .
\end{equation} 
We have the following definitions: Let $[\ ]_r$ denote renormalization
of the operator $J$.  Define $V^{(n)}(a,b)$ through the expansion:
\begin{equation}
 \left[ \exp\left(\int_a^b dr\, \lambda J(r) \right) \right]_r
      = \sum _{n = 0}^\infty \lambda^n \left[ V^{(n)} (a,b) \right]_r \ .
\end{equation}
We can then define $U_\alpha$ through
\begin{equation}
  \langle \phi |U_\alpha\rangle = \sum_{n = 0}^\infty
\lambda^n \langle f\circ \phi(0) \ V^{(n)}(1,n+\alpha) \rangle_{C_{n+\alpha+1}} \ .
\end{equation}
where $C_m$ is the cylinder of width $m$ and $f = \frac{2}{\pi}
\arctan(z)$.  The state $U$ used above is defined to be $U = U_0$.  We
define, also, $A_{L,\alpha}$ and $A_{R,\alpha}$ through the relation
\begin{equation}
  Q_B U_{\alpha} = A_{L,\alpha} - A_{R,\alpha} \ ,
\end{equation}
where we split up the terms on the right hand side by saying that
$A_{L,\alpha}$ is all the terms with ghosts on the left side of the
worldsheet and $A_{R,\alpha}$ is all the terms with ghosts on the
right.  This definition is does not determine the $\lambda^1$ term in general, 
so we supplement it with
\begin{align}
   \langle \phi| A_{L,\alpha} \rangle &= \lambda \langle f\circ \phi(0) \
   cJ(1+\alpha) \rangle_{C_{2+\alpha}} + \mathcal{O} (\lambda^2) \ ,
\\
   \langle \phi| A_{R,\alpha} \rangle &= \lambda \langle f\circ \phi(0) \
   cJ(1) \rangle_{C_{2+\alpha}} + \mathcal{O} (\lambda^2) \ .
\end{align}
As a final definition, we set $A_{L,0} = A_L$ and $A_{R,0} = A_R$.

These states obey the important identities proved in \cite{Kiermaier:2007vu},
\begin{equation}
  U_\alpha *U^{-1} U_\beta = U_{\alpha + \beta} \ , \qquad 
  A_{L,\alpha} *U^{-1} * U_\beta = A_{L,\alpha+\beta}
 \ , 
\qquad 
  U_\alpha * U^{-1} * A_{R,\beta} = A_{R,\alpha+\beta} \ .
\end{equation}

\subsubsection{Computation of $P_\Upsilon$}

We have
\begin{equation}
  \Omega_{\Upsilon} = |0\rangle -\{ A, \Upsilon_0 \} \ .
\end{equation}

Computing $\lim_{N \to \infty} \Omega^N_{\Upsilon}$ directly appears to be an
almost impossible task, though it is easy to perform to low orders in
$\lambda$.  Instead, as has often been the case when working with the
Kiermaier-Okawa solution, it is best to start by working with
${\Upsilon}_{L,R}$ and then understand how the answers change
under a gauge transformation.

Hence, we start instead with
\begin{equation}
  \Omega_{{\Upsilon}_L} = |0\rangle -\{ A, \Upsilon_{0L} \} \ ,
\end{equation}
and consider
\begin{equation} \label{OmegaProd}
  \lim_{N\to \infty} (\Omega_{{\Upsilon}_{L}})^N \ .
\end{equation}
Following the same arguments from the previous section, this is given by (keeping track of only
the left half of the string)
\begin{equation}
  \frac{1}{1 -Z* B_1 A_{L} * U^{-1}} *|\infty\rangle \ .
\end{equation}
Where $Z = \int_0^\infty dr \,W_r$.
We now claim that this yields the same (left half) of the projector
from the previous section, $P_{{\Upsilon}_{L}} \overset{\text{left-half}}{=}
P_{{\Theta}}$, which, in the notation of \cite{Kiermaier:2007vu} is $U_{\infty}$  It turns out to be easiest to show this by
induction in powers of $\lambda$.  At lowest order in $\lambda$,
${\Upsilon}_L = {\Theta} + \mathcal{O}(\lambda)$, so the base case is immediate.  Now
suppose that we have checked the answer to
$\mathcal{O}(\lambda^{N-1})$. It follows that
\begin{equation}
   \sum_{m =0}^{N-1} (Z *B_1A_L *U^{-1}) *|\infty\rangle =  U_{\infty} + \mathcal{O}(\lambda^N) \ .
\end{equation}
We then check that
\begin{multline}
   \sum_{m =0}^{N} (Z *B_1A_L *U^{-1}) *|\infty\rangle
 = |\infty\rangle + (Z *B_1A_L *U^{-1}) \sum_{m =0}^{N-1} (Z *B_1A_L *U^{-1}) *|\infty\rangle
\\
 = |\infty\rangle + (Z *B_1A_L *U^{-1}) * U_{\infty} +\mathcal{O}(\lambda^{N+1})
 = |\infty\rangle + Z * B_1 A_{L,\infty}+\mathcal{O}(\lambda^{N+1}) \ .
\end{multline}
It is now easy to check from the definitions that
\begin{equation}
  |\infty\rangle + Z * B_1 A_{L,\infty}  = U_{\infty} \ ,
\end{equation}
which proves the inductive step.  Hence we have found,
\begin{equation}
  P_{{\Upsilon}_L} \overset{\text{left-half}}{=} P_{{\Theta}} = U_{\infty}
\end{equation}
Similarly, we would find
\begin{equation}
  P_{{\Upsilon}_R} \overset{\text{right-half}}{=} P_{{\Theta}} = U_{\infty} \ .
\end{equation}
A few comments are in order.  Focusing instead on the right half of
$P_{\Upsilon_L}$ does yield the right half of $U_{\infty}$.  This is
related to the fact that ${\Upsilon}_L$ is {\em not} a real string
field.  Furthermore, since we are dealing with a deformation $J$ which
has a non-trivial OPE with itself, the product $U_{\infty}*U_{\infty}
\ne U_{\infty}$, and is divergent.  We will see shortly that the full
projector $P_{\Upsilon_L}$ does have a finite product with itself.  It
is convenient, however, to first compute the left half of the
projector for the full solution, $P_{\Upsilon}$ (which is our main
interest at any rate).

Recalling that ${\Upsilon}$ is a gauge transformation of ${\Upsilon}_L$,
we consider how a general projector $P_{{\Phi}}$ changes under a gauge transformation.
Note that we can write
\begin{equation}
  P_{{\Phi}} = \lim_{N\to \infty} \left(1 - Q_\Phi A \right)^N \ .
\end{equation}
Under a gauge transformation $\Phi \to \Phi + Q_B \Lambda +
\{\Phi,\Lambda\}$, this becomes (keeping terms at lowest order in
$\Lambda$)
\begin{multline}
   \lim_{N\to \infty} \left(1 - Q_\Phi A  +[\Lambda,Q_{\Phi} A] -Q_{\Phi} [\Lambda,A]\right)^N
\\
 = P_{{\Phi}} - [\Lambda, P_{{\Phi}}]
+ \frac{1}{1 + Q_{\Phi}A}* Q_{\Phi} ([\Lambda,A]) * P_{{\Phi}}
+ P_{{\Phi}}*Q_{\Phi} ([\Lambda,A]) * \frac{1}{1 + Q_{\Phi}A}
\\
 = 
 P_{{\Phi}} - [\Lambda, P_{{\Phi}}]
- \frac{1}{1 + Q_{\Phi}A}* Q_{\Phi} (A*\Lambda * P_{{\Phi}})
+ Q_{\Phi} (P_{{\Phi}}*\Lambda*A) * \frac{1}{1 + Q_{\Phi}A} \ ,
\end{multline}
where we have used that $A*P_{{\Phi}} = P_{{\Phi}}*A = 0$ and
$Q_{\Phi} P_{{\Phi}} = 0$.  For the case we are considering, we
have that $\Lambda = f(U)$ and hence $A*\Lambda*P_{{\Phi}} =
P_{{\Phi}}*\Lambda*A = 0$ so we have the simple change,
\begin{equation} \label{PChange}
  P_{\Phi+Q_B\Lambda + \{\Phi,\Lambda\}} = P_{\Phi} -[\Lambda,P_{{\Phi}}] + \mathcal{O}(\Lambda^2)  \ .
\end{equation}
Integrating the infinitesimal form (\ref{PChange}), we find for the complete
solution ${\Upsilon}$, the {\em left half} of the projector is given by
\begin{equation} \label{PL}
  P_{{\Upsilon}} = \frac{1}{\sqrt{U}} P_{\Upsilon_L} \sqrt{U} 
 \overset{\text{left half}}{=} \frac{1}{\sqrt{U}} U_{\infty}\ .
\end{equation}
This does not give the correct right half since $U_{\infty}$ is only
what we get on the left half of $P_{{\Upsilon}_L}$.  Redoing the entire computation
using ${\Upsilon}_R$, we find
\begin{equation}\label{PR}
  P_{{\Upsilon}} =  \sqrt{U} P_{\Upsilon_R} \frac{1}{\sqrt{U}}
 \overset{\text{right half}}{=}   U_{\infty} \frac{1}{\sqrt{U}} \ .
\end{equation}
In total, we must have
\begin{equation}
  P_{{\Upsilon}} = \frac{1}{\sqrt{U}} U_{\infty} \frac{1}{\sqrt{U}}  \ .
\end{equation}
In fact, this answer could have been guessed from the discussion in
\cite{Kiermaier:2007vu} since $ P_{{\Upsilon}}$ is the natural analogue of the
sliver for the marginal solution.  It has a finite product with itself,
\begin{equation}
 P_{{\Upsilon}} * P_{{\Upsilon}} = P_{{\Upsilon}} \ .
\end{equation}
which follows from the general formula,
\begin{equation}
  U_r * U^{-1} * U_s = U_{r+s} \ .
\end{equation}
It also is of the same form as in the trivial OPE case $
P_{{\Upsilon}} \sim \Xi* |\infty_{\lambda J} \rangle *\Xi^{\ddagger}$.  In this
case, the fields $\Xi$ are significantly more complicated to account
for the divergences when operators collide.  

Note that, using (\ref{PL}) and (\ref{PR}), we see that
\begin{equation}
  P_{\Upsilon_L} = U_\infty* U^{-1} \ , \qquad P_{\Upsilon_R} = U^{-1} U_\infty \ .
\end{equation}
These fields satisfy, $ P_{\Upsilon_{L,R}}* P_{\Upsilon_{L,R}}
=P_{\Upsilon_{L,R}}$.  However,
$P_{\Upsilon_{R}}*P_{\Upsilon_{L}}$ is divergent.  

We close this section with a few remarks.  We find it curious that there appear to be two candidates for
the analogues of wedge states of the deformed theory.  On the one hand, as proposed in
\cite{Kiermaier:2007vu}, the states
\begin{equation}
  \widehat{W}_\alpha = \frac{1}{\sqrt{U}} U_\alpha \frac{1}{\sqrt{U}} \ ,
\end{equation}
satisfy $\widehat{W}_\alpha * \widehat{W}_\beta = \widehat{W}_{\alpha
+\beta}$.  At least for {\em integer} values of  $\alpha$, we can also define
\begin{equation}
  \Omega_\alpha = \Omega_{\Upsilon}^\alpha \ ,
\end{equation}
which trivially satisfies $\Omega_\alpha *\Omega_\alpha =\Omega_{\alpha +\beta}$.
Both ``wedges'' are annihilated by $Q_\Upsilon$ and limit to $P_{\Upsilon}$ as $\alpha \to \infty$.  

Finding the cohomology of $Q_{\Upsilon}$ using $P_{\Upsilon}$ is more
subtle than in the trivial OPE case because the projector only has
boundary conditions associated with $\Upsilon$ {\em near} the midpoint. However, to any order in $\lambda$
there is a finite width region around the midpoint with modified boundary conditions to that order.  Hence, assuming
the cohomology can be computed perturbatively in $\lambda$ one can construct representatives in an identical
way to the trivial OPE case.

\section{A possible method for constructing new solutions} \label{NewSolutions}

In this section we would like to propose a new method for constructing non-trivial string field theory
solutions.  As an application, we will propose a definite solution representing lower dimensional branes.  
Whether or not this technique leads to finite solutions with computable energy remains to be seen.

The idea is to write $\Phi = U^{-1} \mathcal{Q} U$ with $U$ a singular gauge transformation,
keeping as many of the properties of $U$ as possible from the known solutions.
Recall that, for a solution to the equations of motion $\Phi$ we have $U = \nol  f(\Omega^L,\Omega^R)\nor $,
where 
\begin{equation} \label{OmegaDefs}
  \Omega^L = -A*\Phi \ , \qquad  \Omega^R = -\Phi*A \ , \qquad \text{and} \qquad \mathcal{Q}\Phi + \Phi*\Phi = 0\ .
\end{equation}
From our earlier analysis, we found that the gauge transformation $U$ was singular when we could
find a projector $P$ such that  $\Omega^L*P = P$ and $P*\Omega^R = P$.

To find new non-trivial solutions, we relax the conditions  (\ref{OmegaDefs}) and impose only
\begin{equation} \label{OmegaConditions}
\Omega^L = -A*\Gamma \ ,\qquad \Omega^R = -\Gamma*A \ , \qquad \Omega^L*P = P \ , \qquad \text{and} \qquad  P*\Omega^R = P \ .
\end{equation}
where $P$, following the example of the known solutions, is a projector with new boundary conditions at the midpoint and
{\em importantly}, we no longer assume that $\Gamma$ is a solution to the equations of motion, only that $\Gamma^{\ddagger} = \Gamma$.
Finding an appropriate projector $P$ is, in many cases, not difficult.  Solving the conditions (\ref{OmegaConditions}) could, in general be challenging.

We can then conjecture that if we  take $\Pi = U^{-1} \mathcal{Q} U$ with  $U = \nol  f(\Omega^L,\Omega^R)\nor $ and
$\Omega^{L,R}$ satisfying  (\ref{OmegaConditions}), then $\Pi$ will be solution to the equations of motion
corresponding to BCFT given by the boundary conditions near the midpoint of $P$.

Note that $\Pi$ will be a real solution ($\Pi$ = $\Pi^{\ddagger}$) which is not pure-gauge (since the gauge transformation is singular).
It is possible (if not likely for generic $\Omega^{L,R}$), however, that $\Pi$ diverges in the level expansion.

Consider also
\begin{equation}
  A* \Pi *A = A*U^{-1}*\mathcal{Q}U*A  \ .
\end{equation}
Using the explicit form for $U^{-1}$, this reduces to
\begin{multline} \label{APhiA}
  A* \nol f(\Omega^R,\Omega^L)\nor *\mathcal{Q}U*A
   = A* \nol f (\Omega^R,0)\nor  *  \mathcal{Q}U*A 
   \\
   = A*\sqrt{1-\Omega^R}*\mathcal{Q}U*A = \sqrt{1-\Omega^L}*A*\mathcal{Q}U*A \ .
\end{multline}
Now, when we look inside $A*\mathcal{Q}U*A$, we see that only the terms in $\mathcal{Q} U$ with no $A$'s on either the right or the left are non-vanishing.  For terms with both $\Omega^L$s and $\Omega^R$s this is impossible.  For terms with only $\Omega^L$s, the $\mathcal{Q}$ must hit the first $A$ which is then replaced by the left $A$ in $A*\mathcal{Q}U*A$.  The same thing happens for terms with only $\Omega^R$s except that there is an extra minus.  In total, one gets
\begin{equation}
  A*\mathcal{Q}U*A = (\sqrt{1-\Omega^L} - 1)*A - A*\left(\frac{1}{\sqrt{1-\Omega^R} }- 1\right)
   = -\frac{\Omega^L}{\sqrt{1-\Omega^L}}*A  \ .
\end{equation}
Examining (\ref{APhiA}), we see that
\begin{equation}\label{PsiGammaRule}
  A*\Pi*A = -\Omega^L*A = A*\Gamma*A \ .
\end{equation}
This identity may seem unexpected.  However, it can be explained as follows:  If $\Gamma$ satisfied the equations of motion, then we would have $\Gamma = \Pi$ (by construction) and the identity would be trivial.  The fact that $\Gamma$ doesn't satisfy the equations of motion only enters in terms where $\mathcal{Q}$ hits $\Gamma$.  Because of the $A$'s on the left and right, these terms do not contribute.

Using (\ref{PsiGammaRule}), we find the identities,
\begin{equation}
  (1+A*\Pi)*P = P*(1+\Pi*A) = 0 \ ,
\end{equation}
as well as
\begin{equation}
  \lim_{N\to\infty} (-\{A,\Pi\})^N =  \lim_{N\to\infty} (-\{A,\Gamma\})^N = P \ .
\end{equation}
assuming that we have picked $\Gamma$ appropriately.  This implies that we can use $P$ as a basis for elements of the cohomology around the $\Pi$ vacuum just as we did in the case of the known solutions.  At the very least, solutions arising from this technique will have the correct spectrum provided that they exist at all (i.e. don't diverge in the level expansion).

\subsection{Lower dimensional branes}

Here we give a concrete example:  a lower dimensional brane solution.  This solution has not been checked in the level expansion.  We hope to do so in the near future, but at present the solution is only a proposal.  Even if this particular guess of the form of the solution turns out to be singular, we hope that the projector and $\Omega$'s described here will be useful in later study.

We begin first by constructing the characteristic projector $P$ which will have Dirichlet boundary conditions at the midpoint and
Neumann boundary conditions near its endpoints.  To construct such a projector it is useful to recall that there is an family of boundary
conditions which interpolate between Dirichlet and Neumann.  These boundary conditions are {\em not} conformal. If they were, we could construct a marginal solution.

Let $y$ be a coordinate on the boundary and
\begin{equation}
  \mathcal{T}_u (y) = \frac{u}{8\pi} :X^2:(y) + \frac{1}{2\pi} ((\gamma -1)u+u\log u) \ ,
\end{equation}
where $\gamma$ is the Euler-Mascheroni constant and $X$ is one of the spacetime coordinates.
Then inserting
\begin{equation}\label{BoundaryDeformationByT}
  \exp\left( -\int dy\, \mathcal{T}_u (y)\right)
\end{equation}
into a correlator with Neumann boundary conditions gives a non-conformal (but Gaussian) modification of the Neumann conditions.  The key property of $\mathcal{T}_u$ is that sending $u\to \infty$ gives Dirichlet boundary conditions for the $X$ coordinate.  This occurs because the $X^2$ term constrains the endpoints of the string to be near $X =0$.   The $u\log u$ piece in $\mathcal{T}_u$ can be determined by conformal invariance.  The $(\gamma -1)u$ term must be determined by computing the partition function with the $\mathcal{T}_u$ deformation and ensuring that it has a finite limit as $u \to \infty$.  This computation is performed in appendix \ref{DefConst}.
It is important to note that, although there are logarithmic singularities when two $\mathcal{T}_u$'s approach one another, these singularities are integrable, and therefore (\ref{BoundaryDeformationByT}) does not require any regularization.

Although the boundary condition is not conformal, it does have a simple transformation under conformal transformations.  One finds
\begin{equation}
  f \circ \int_a^b dy \,\mathcal{T}_u(y) = \int_{f(a)}^{f(b)} d\tilde{y}\, \mathcal{T}_{u \partial_{\tilde{y}} f^{-1}(\tilde{y})} (\tilde{y}) \ ,
\end{equation}
which can be verified from the transformation law,
\begin{equation}
  f\circ :X^2:(y) = :X^2:(f(y)) - 2 \log |f'(y)|^2  \ .
\end{equation}
Now, consider $P$ defined by
\begin{equation}
  \langle \phi | P\rangle = \left\langle  f\circ \phi  \exp\left( - \int_{1}^\infty dy\,\mathcal{T}_u(y) \right)
  \exp\left(-\int_{-\infty}^{-1}dy\,  \mathcal{T}_u (y)\right)
  \right\rangle_{\text{UHP}} \ .
\end{equation}
This defines $P$ to be a sliver-like surface state with $\mathcal{T}_u$-deformed boundary conditions along its boundary (except near the endpoints where the boundary is undeformed).

Consider the behavior of $P$ near its midpoint.  To do so, map $z \to w(z) =-1/z$.   Noting that $w'(z) = 1/z^2$, near the midpoint $\tilde{y} = w = 0$, the boundary is deformed by
\begin{equation}
  \exp\left(- \int d\tilde{y} \mathcal{T}_{u/{\tilde{y}}^2} (\tilde{y}) \right) \ .
\end{equation}
Hence, near the midpoint of $P$, the deformation parameter is sent to infinity and the boundary conditions become Dirichlet.
One can now ask whether $P*P = P$.  We will verify this indirectly in a moment.  Note that our projector depends on $u$.  This
will remain as a gauge parameter in the solution\footnote{This gauge parameter is similar to the constant $b$ in the analogous  vacuum string field theory solution \cite{Rastelli:2001jb}.  Taking $u$ small, the solution limits to the perturbative vacuum near the origin.  Taking $u$ large, the solution appears to limit to the tachyon vacuum (except presumably near the origin) using arguments similar to those in \cite{Ellwood:2007xr}.  We expect, therefore, that $u$ determines the width of the lump with $u \to \infty$ being the limit of zero width.}.  Note that by changing the width of the non-deformed edges of $P$ and reparametrizing
is enough to change $u$ to any finite value.

We would now like to construct $\Omega^{L,R}$ satisfying (\ref{OmegaConditions}).  To do so, compare the form of $P$ with $P_\Theta$ given in equation (\ref{TrivialOPEP}).  These two projectors are the same except that in $P$ we have replaced $\lambda J \to -\mathcal{T}_u$.  It follows that if we replace every occurrence of $\lambda J$ in the string field $\Theta$ by $\mathcal{T}_u$, the resulting string field,
\begin{equation}
  \Gamma = -\Psi +W_{1/2}* \frac{1}{1+c \mathcal{T}_u*A} *c \mathcal{T}_u* W_{1/2} \ ,
\end{equation}
can be used to construct
\begin{equation}
  \Omega^L = -A*\Gamma \ , \qquad \Omega^R  = -\Gamma*A  \ ,
\end{equation}
which automatically satisfy (\ref{OmegaConditions}).   Note that, as was true in the marginal case, 
\begin{equation}
  \lim_{N\to \infty} (\Omega^L+\Omega^R)^N = P \ .
\end{equation}
This is enough to guarantee that $P*P = P$.  Note, however, that it does not rule out the possibility that $P = 0$ or $P$ diverges.  Ensuring a finite $P$ requires a careful choice of the normal ordering constant in $\mathcal{T}_u$ which we discuss in appendix \ref{DefConst}.

A few comments are in order.  First, $\Gamma$ is a finite string field.  Although there are collisions between $\mathcal{T}_u$, they are integrable.  Second, $\Gamma$ is not a solution to the equations of motion.  This happens because $Q_B(c \mathcal{T}_u) \ne 0$.

However, although $\Gamma$ is not a solution, we expect that
\begin{equation}
  \Pi = \nol f(\Omega^R,\Omega^L)\nor  \mathcal{Q}( \nol f(\Omega^L,\Omega^R)\nor )
\end{equation}
will satisfy the equations of motion since it is in pure-gauge form.  It remains to be seen, however,  whether $\Pi$ is a finite solution, and if so, whether it satisfies the equations of motion when contracted with itself.
At the very least, if it is finite, we expect using the cohomology arguments described above, it will have the correct cohomology.  Clearly, though, further tests are required before anything more concrete can be said.

\section{An equivalence between two OSFTs} \label{s:OSFTequivalence}

In this last section we explain a simple picture that demonstrates that the OSFT around the shifted vacuum is
equivalent to the OSFT given by changing the BCFT used to define the
theory.  For the marginal solution with trivial OPE, this was shown in
\cite{Ellwood:2007xr} in a very different way.  For infinitesimal
marginal transformations, another demonstration was given in
\cite{Sen:1990hh,Sen:1990na,Sen:1992pw,Sen:1993mh}.

Our construction is rather formal (in that it involves projector-like states), and relies heavily on the assumption that characteristic projector has boundary conditions near its midpoint corresponding to the deformed BCFT.  Nonetheless, we believe it yields some intuition about
characteristic projectors and solutions that cannot be seen in another way.

\begin{figure}
\centerline{
\begin{picture}(407,163)(0,-10)%
\includegraphics{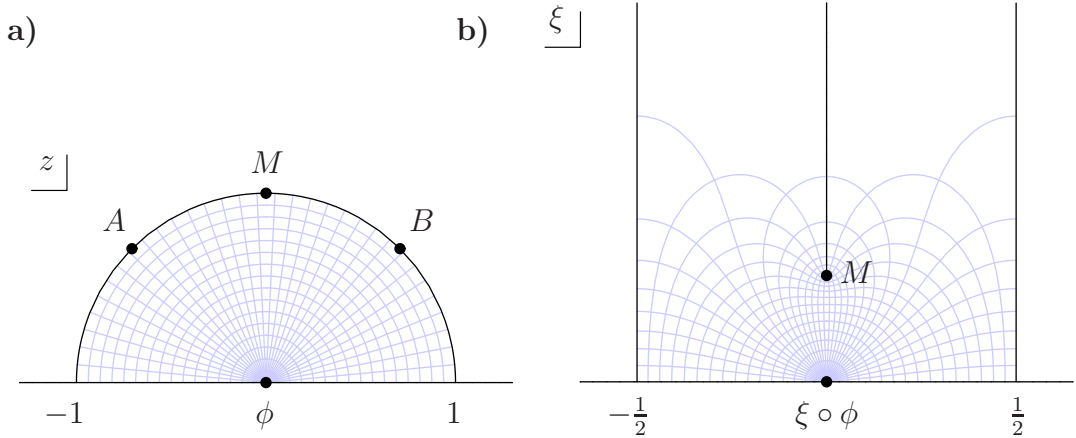}%
\end{picture}%
\begin{picture}(0,0)(407,-10)
\put(0,135){{\bf a)}}
\put(94,-10){$\phi$}
\put(92,85){$M$}
\put(36,63){$A$}
\put(152,63){$B$}
\put(12,86){$z$}
\put(166,-10){$1$}
\put(14,-10){$-1$}
\put(170,135){{\bf b)}}
\put(205,140){{$\xi$}}
\put(315,43){{$M$}}
\put(227,-10){{$-\frac{1}{2}$}}
\put(298,-10){{$\xi \circ \phi$}}
\put(379,-10){{$\frac{1}{2}$}}
\end{picture}}
\caption{The method for constructing a state in which the middle of
  the string is identity-like is illustrated.  In a), we show the
  coordinate patch in the $z$-coordinate in the UHP for a test field
  $\phi$.  In b), the coordinate patch is mapped via $\xi(z)$ to the
  $\xi$-coordinate.  The arcs $M$-$A$ and $M$-$B$ in the
  $z$-coordinate are identified and mapped to the vertical line above
  $M$.  The points $A$ and $B$ themselves are mapped to $i\infty$.
  The arcs from $A$ to $-1$ and $B$ to $1$ are mapped to the vertical
  lines above $-1/2$ and $1/2$.
 \label{ximap}}
\end{figure}

We begin by defining a $*$-algebra homomorphism, which can be thought
of as a generalization of a reparametrization.  Imagine that a state
$\Psi$ is defined in the standard $f(z) = \frac{2}{\pi}\arctan (z)$
coordinates.  In these coordinates, the coordinate patch is given by
the region of the UHP with $-\frac{1}{2} \le \Re (z) \le \frac{1}{2}$.

Instead of using $f(z)$, we can use any other map, $\xi(z)$, which
maps the coordinate patch to the same region of the upper half plane.
We can compute a new state $\rho(\Psi)$ by
replacing the coordinate patch found using the map $f(z)$ with the
modified coordinate patch $\xi(z)$.  Algebraically, suppose
\begin{equation}
  \langle \phi |\Psi \rangle = \langle f\circ \phi(0) \ \mathcal{O}_\Psi(1) \rangle_{C_2} \ ,
\end{equation}
where $\mathcal{O}_\Psi$ is some operator. Then
\begin{equation}
   \langle \phi |\rho(\Psi) \rangle = \langle \xi\circ \phi(0) \ \mathcal{O}_\Psi(1) \rangle_{C_2}
\end{equation}

If $f^{-1}\circ \xi$ is an
automorphism of the unit disk which preserves $z\in \{-1,1,i\}$, this
procedure simply defines a reparametrization \cite{Okawa:2006sn}.  
More generally however, we may consider cases in which $\rho$ is {\em
not} simply a reparametrization.  In particular, we are interested in
the case,
\begin{equation}
 \xi(z) = \tfrac{i}{\pi} \left(\arctan(e^{-i\theta} z) -\arctan(e^{i\theta} z) \right) \ .
\end{equation}
As illustrated in figure \ref{ximap}, an important feature of the map
$\xi$ is that for $\theta \le \arg(z) \le \pi - \theta$, the left and
right halves of the coordinate patch are glued together.  The variable
$\theta$ therefore determines to what extent $\rho(\Psi)$ is identity-like.
Sending $\theta \to \pi/2$, one finds $\xi(z) \to f(z)$, so
$\rho(\Psi) = \Psi$, while sending $\theta \to 0$, the entire state
$\Psi$ is squeezed to the ends of the string and $\rho(\Psi)$ becomes
the identity state with a possibly divergent operator insertion on the
boundary.  Different $\theta$ are related to each other by
reparametrization, except $\theta = 0$ and $\theta = \pi/2$.

Because $\rho$ satisfies the identities,
\begin{equation}
  \rho(\Psi_1)*\rho(\Psi_2) = \rho(\Psi_1 * \Psi_2) \ ,
\qquad  Q_B \rho (\Psi) = \rho(Q_B\Psi)
\end{equation}
it follows that if $\Phi$ is a solution, then $\Phi' = \rho(\Phi)$ is
also a solution.  Since $\Phi'$ can be limited to $\Phi$ by a sequence
of reparametrizations (sending $\theta \to \pi/2$), we expect $\Phi'$
is gauge equivalent to $\Phi$.  However, $\Phi'$ turns out to be very
convenient for cohomology computations because its left and right
action by $*$-multiplication leaves an entire region near the midpoint
alone.

\begin{figure}
\centerline{
\begin{picture}(384,163)(0,-10)
\includegraphics{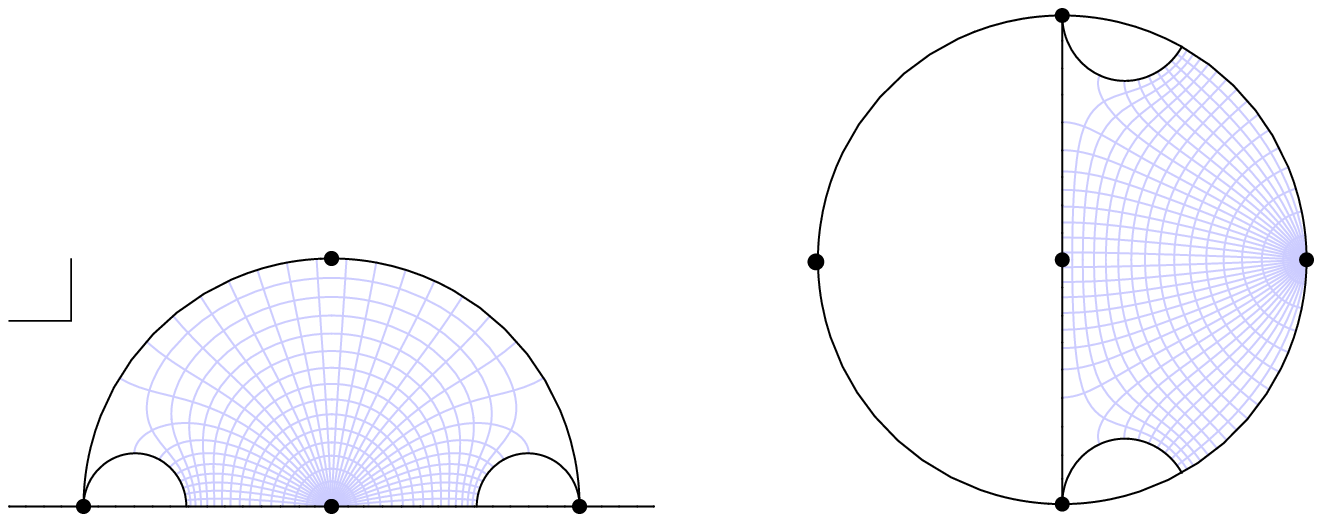}%
\end{picture}%
\begin{picture}(0,0)(384,-10)
\put(0,135){{\bf a)}}
\put(87,-10){$\eta \circ \phi$}
\put(22,-10){$A$}
\put(165,-10){$B$}
\put(93,85){$M$}
\put(10,66){$\eta$}
\put(220,135){{\bf b)}}
\put(304,-10){$A$}
\put(304,155){$B$}
\put(387,73){$h\circ \phi$}
\put(290,73){$M$}
\put(220,73){$\Sigma$}
\end{picture}}
\caption{In a) it is shown how the entire geometry from figure
\ref{ximap}b can be mapped into the upper unit disk by the inverse of
the map associated with the identity state, $\eta(\xi)$.  This
procedure unsews the segments $A$-$M$ and $B$-$M$ allowing one to
attach a standard $SL_2(\mathbb{R})$ vacuum.  The two small
semi-circles above the boundary are the edges of the left and right
half of the projector $P$.  In b) the same geometry is shown in
disk coordinates.  Additionally, a possible operator insertion
$\Sigma$ is shown.
 \label{Unfolding}}
\end{figure}

To complete the discussion, we need one more ingredient: Consider the case
when $\Psi$ is a sliver-like projector, $P$.  In this case, in figure \ref{ximap}b, the
entire UHP is filled with worldsheet on the left and right of the coordinate patch.
Rescaling everything by $\xi \to w(\xi) = \lambda^{-1} \xi$, where $\lambda$ is the height
of the midpoint $M$,
\begin{equation}
  \lambda = \xi(i) = -\tfrac{i}{\pi} \log\left(i\, \frac{e^{i\theta} - i}{e^{i\theta}+i} \right)  \ ,
\end{equation}
we can apply the map, 
\begin{equation}
 \eta(w) =  i \,\,\frac{1-\sqrt{\frac{i-w}{i+w}}}{1+\sqrt{\frac{i-w}{i+w}}} \ ,
\end{equation}
which is the inverse of the familiar map associated with the identity state
\begin{equation}
  w(\eta) = \frac{2 \eta}{1-\eta^2} \ .
\end{equation}
Because of the branch cut extending from $i$ to infinity, $\eta(\xi)$
unsews the left and right halves of the coordinate patch that were
glued together by $\xi(z)$ and sends them to the boundary of the unit disk.
The result is shown in figure \ref{Unfolding}.

We can now fill the rest of the upper half plane with worldsheet,
effectively attaching an $SL_2(\mathbb{R})$ vacuum.  It will be
convenient to include, as well, a vertex operator $\Sigma$ as shown in
figure \ref{Unfolding}b.  We will think of this construction as
defining a map from a state $\Sigma = \Sigma(0)|0\rangle$ to a state
$\tau_P(\Sigma)$, which is a function
of $\Sigma$ and the projector, $P$.

Now, suppose that $P = P_{{\Phi}}$ is one of the characteristic projectors
found earlier.  An important property of the states $P_{{\Phi}}$
we found in our discussion is that near the midpoint, the boundary condition limits to the BCFT${}_\Phi$ associated with the state
${\Phi}$.  However, examining figure \ref{Unfolding}a and b, we
see that the midpoint of $P_{{\Phi}}$ is glued to the edge of the
state $\Sigma$.  Hence, in order to have a finite state
$\tau_{P_{{\Phi}}}(\Sigma)$, we must enforce the boundary
condition BCFT${}_\Phi$ on the edge of $\Sigma$.  It follows that we
can think of $\tau_{P_{{\Phi}}}$ as defining a map from the OSFT
defined with boundary condition BCFT${}_\Phi$ into the Fock space of
our original OSFT.

Notice that
\begin{equation} \label{tauAsHomomorphism}
  \tau_P(\Sigma_1)*\tau_P(\Sigma_2) = \tau_P(\Sigma_1*\Sigma_2) \ ,
\end{equation}
which follows from $P*P = P$ and the usual gluing rules of star multiplication.
One can also see that
\begin{equation}\label{Qonrho}
  Q_B \tau_P(\Sigma) = \tau_{Q_B P}(\Sigma)  +\tau_P(Q_B \Sigma)  \ .
\end{equation}
Finally, one has
\begin{equation}\label{taurhostar}
  \rho(\Psi) * \tau_P(\Sigma) = \tau_{\Psi*P} (\Sigma) \ ,
\qquad \tau_P(\Sigma)*\rho(\Psi)  = 
(-1)^{\Sigma \Psi} \tau_{P*\Psi} (\Sigma) \ .
\end{equation}
Combining (\ref{Qonrho}) and (\ref{taurhostar}) we see that
\begin{equation}
  Q_{\rho(\Phi)} \tau_{P_\Phi}(\Sigma) = \tau_{Q_\Phi P_\Phi}(\Sigma) + 
\tau_{P_{\Phi}}(Q_B \Sigma)  = \tau_{P_{\Phi}}(Q_B \Sigma) \ .
\end{equation}
So that the complicated operator $Q_{\rho(\Phi)}$ on $\tau_{P_\Phi} (\Sigma)$
just reduces to the ordinary BRST operator (in BCFT${}_{\Phi}$) acting
on $\Sigma$.

Noting (\ref{tauAsHomomorphism}) implies that $\tau$ is a $*$-algebra homomorphism,
we see that we have a complete map from the OSFT defined around BCFT${}_\Phi$
into our original OSFT:  Any computation in OSFT${}_\Phi$ can be performed 
by mapping 
\begin{equation}
\Sigma \to \tau_{P_{\Phi}}(\Sigma) \ , \qquad Q_B \to Q_{\rho(\Phi)} \ .
\end{equation}
In this map, the state $P_{\Phi}$ acts like a boundary condition
changing operator, while $\Phi$ cancels out the extra terms that come
from $Q_B$ acting on $P_\Phi$.  Note that the vertex operator in $\Sigma$ is not required to
be a weight $(0,0)$ primary, since it is not inserted near the string
midpoint.

\section*{Acknowledgments}

I would like to thank S.~Das, W.~Merrel and A.~Shapere for
discussions and T.~Erler and M.~Schnabl for comments on the draft.  I would also like to thank the Institute of Physics at the Czech Academy of Sciences where part of this work was completed.  This work was supported by Department of Energy Grant
No. DE-FG01-00ER45832.

\appendix
\section{Non-projector-like representatives of cohomology for the trivial OPE case} \label{s:NonProjCohomology}

In this appendix we give non-projector-like representatives of the cohomology for the BRST operator $Q_{\Theta_0}$
where $\Theta_0$ is the marginal solution with trivial OPE.  A special class of elements of the cohomology was found in \cite{Schnabl:2007az} and our construction uses similar techniques.

Let $\Sigma$ be an element of the space of states of the deformed CFT corresponding to 
the solution $\Phi$.  Consider the state,
\begin{equation}
  \pi(\Sigma)= W_{1/2}*\frac{1}{1-cJ *A} *\Sigma * \frac{1}{1-A*cJ} *W_{1/2}  \ .
\end{equation}
Since $\Sigma$ is in the CFT corresponding to $\Phi$, we have
\begin{equation}
  Q_B \Sigma = Q^\Phi_B \Sigma + [cJ,\Sigma] \ ,
\end{equation}
where $Q^\Phi_B$ is the BRST operator of the CFT corresponding to $\Phi$ (which does not include the contribution from $Q_B$ hitting the boundary condition changing operators) and the second term arises from the BRST
operator $Q_B$ acting on the boundary condition changing operators.

To proceed, we compute
\begin{multline}
  Q_B \pi(\Sigma) = 
  W_{1/2} * \frac{1}{1 - cJ *A}* cJ ( |0\rangle-1) *\frac{1}{1-cJ*A} * \Sigma * \frac{1}{1-A*cJ} *W_{1/2}
  \\
  + W_{1/2} * \frac{1}{1 - cJ *A}  * (Q_B^\Phi \Sigma + [cJ,\Sigma]) * \frac{1}{1-A*cJ} *W_{1/2}
  \\
  + (-1)^\Sigma  W_{1/2}*\frac{1}{1-cJ *A} *\Sigma * \frac{1}{1-A*cJ}  * (1-|0\rangle)*cJ*  \frac{1}{1-A*cJ} *W_{1/2} \ ,
\end{multline}
which reduces to
\begin{equation}
  Q_B \pi(\Sigma) = - \{ \Theta_0, \pi(\Sigma)\} + \pi(Q_B^\Phi \Sigma) \ .
\end{equation}
Hence, one has the convenient identity,
\begin{equation} \label{Qmorphism}
  Q_\Phi \pi (\Sigma) = \pi(Q_B^\Phi(\Sigma)) \ ,
\end{equation}
mapping the cohomology problem of the CFT deformed by $J$ into SFT around the solution $\Phi$.  This construction fails in the non-trivial OPE case due to operator collisions and no simple modifications of the map $\pi$ appear to work. Note that
\begin{equation}
 \pi(\Sigma_1 *\Sigma_2) \ne \pi(\Sigma_1)*\pi(\Sigma_2) \ ,
\end{equation}
so that we do {\em not} have a $*$-algebra homomorphism.   It does not appear to be straightforward to construct a completely non-singular homomorphism satisfying (\ref{Qmorphism}).

\section{Another form for the manifestly real pure-gauge transformation} \label{PGF}

Upon examination of a draft of this paper, it was suggested by Erler \cite{ErlerConversation} based on his work \cite{Erler:2007rh} that the gauge transformation (\ref{Uformula}) might take the alternate form,
\begin{equation}
  \frac{1}{\sqrt{(1-\Omega^L)*(1-\Omega^R)} }*(1-\Omega^L) \ .
\end{equation}
In this appendix, we give a proof of this suggestion.

We begin with some formulas for the algebra generated by $\Omega^{L,R}$ with the relation $\Omega^R \Omega^L = 0$.
Let $f(x,y)$ and $g(x,y)$ be ordinary functions of two variables which are analytic around $x = 0$ and $y = 0$.  We find the
multiplication law,
\begin{equation} \label{MultiplicationLaw}
  \nol f(\Omega^L,\Omega^R) \nor \, \nol g(\Omega^L,\Omega^R) \nor 
  = \nol f(\Omega^L,0) g(\Omega^L,\Omega^R) + f(\Omega^L,\Omega^R) g(0,\Omega^R) 
   - f(\Omega^L,0) g(0,\Omega^R)\nor \ .
\end{equation}
Additionally, using the multiplication law repeatedly, we have the composition law,
\begin{multline} \label{CompositionLaw}
    h(\nol f(\Omega^L,\Omega^R)\nor)
     = 
     \\
     \nol \frac{1}{f(\Omega^L,0) - f(0,\Omega^R)}
      \Bigl[
      f(\Omega^L,\Omega^R) (h(f(\Omega^L,0)) - h(f(0,\Omega^R)))
      \\
       -  (h(f(\Omega^L,0)) f(0,\Omega^R) - f(\Omega^L,0)h(f(0,\Omega^R)) 
      \Bigr] \nor \ ,
\end{multline}
where $h(x)$ is an analytic function around $x = 0$.

Using
\begin{equation}
  (1-\Omega^L)*(1-\Omega^R) = \nol (1-\Omega^L)*(1-\Omega^R) \nor
\end{equation}
we learn from (\ref{CompositionLaw}) that 
\begin{equation} 
  \frac{1}{\sqrt{(1-\Omega^L)*(1-\Omega^R)} }
   = \nol
   \frac{1}{\Omega^L - \Omega^R} \left[ \frac{\Omega^L(1-\Omega^R)}{\sqrt{1-\Omega^L}} - \frac{\Omega^R(1-\Omega^L)}{\sqrt{1-\Omega^R}}\right]
   \nor
\end{equation}
We can then use the multiplication law (\ref{MultiplicationLaw}) to learn that
\begin{equation}
  \frac{1}{\sqrt{(1-\Omega^L)*(1-\Omega^R)} }*(1-\Omega^L) 
   = \nol \frac{1-\Omega^L}{\Omega^L - \Omega^R} \left[ \frac{\Omega^L}{\sqrt{1-\Omega^L}} - \frac{\Omega^R}{\sqrt{1-\Omega^R}} \right] \nor \ ,
\end{equation}
reproducing (\ref{fForm}).

Using the formula for $\mathcal{Q}(\nol f(\Omega^L,\Omega^R) \nor)$ given in (\ref{fEq2}), the reader can verify that
\begin{equation}
  \mathcal{Q}\left(  \frac{1}{\sqrt{(1-\Omega^L)*(1-\Omega^R)} }\right) = 0 \ .
\end{equation}
It follows that
\begin{equation}
   \frac{1}{\sqrt{(1-\Omega^L)*(1-\Omega^R)}} = \mathcal{Q} \left(\frac{1}{\sqrt{1-\Omega^L}}*A \right)
    = \mathcal{Q} \left(A*\frac{1}{\sqrt{1-\Omega^R}} \right) \ .
\end{equation}
These identities allow for an alternate proof that $U^{-1}\mathcal{Q} U = \Phi$.

\section{Normalization of the Gaussian deformation} \label{DefConst}

In this appendix, we explain the our choice 
\begin{equation} \label{boundarydeformation}
  \mathcal{T}_u = \frac{u}{8\pi} :X^2: + \frac{1}{2\pi} ((\gamma - 1)u + u\log u) \ .
\end{equation}
Recall that we have introduced $X^2$ into the boundary action to constrain the endpoints of the string to $X = 0$ in the
$u \to \infty$ limit.  However, different definitions of the operator $X^2$ will lead to a different additive constant term in the boundary action.
The main goal of this appendix is to fix the additive constant, as a function of $u$, so that as $u\to \infty$ one gets Dirichlet boundary conditions.
We can partially fix the coefficients by insisting that $u$ parametrizes an RG flow so that $u \to \infty$ will be a conformal theory in the infrared. The most general boundary action with this property is
\begin{equation} \label{Deformation}
   - \frac{u}{8\pi}\int d\theta  :X^2:- \frac{1}{2\pi} \int d\theta \,u\log u + \frac{A}{2\pi} \int d\theta\, u  \ ,
\end{equation}
where $A$ is an unknown constant.  One can check that rescaling $z \to \lambda z$ takes $u \to \lambda^{-1} u$ as desired.   The question at hand, then, is how to determine $A$.  We do this by studying the partition function as a function of $u$.  In the limit $u\to\infty$ we want to produce Dirichlet boundary conditions, so the partition function should be finite in this limit.  The complete partition function with the deformation  (\ref{Deformation}) was computed by Witten \cite{Witten:1992cr} (up to an overall constant independent of $u$),
\begin{equation}
Z(u) = \sqrt{u} \exp(\gamma u) \Gamma(u) e^{-Au - u\log u} \ .
\end{equation}
Computing the large $u$ limit, one finds
\begin{equation}
  Z(u) \to  \sqrt{2\pi} e^{(\gamma - 1)u-Au}\left( 1+\mathcal{O}(u^{-1}) \right) \ .
\end{equation}
Hence, in order to find a finite partition function in the Dirichlet limit, we take $A = (\gamma -1)$.

Unfortunately, we are not quite done.  In our application of the boundary deformation (\ref{boundarydeformation}), we don't modify the boundary action, but instead insert directly into the correlator $\exp( -\frac{u}{8\pi} \int d\theta :X^2: )$.  Because of the normal ordering, there is potentially a shift in the constant piece relative to Witten's expression.  Happily, as we now describe, this piece is zero:

We begin by computing the disk correlator,
\begin{equation}
 K(u) = \left\langle \exp\left(-\frac{u}{8\pi} \int d\theta :X^2:(e^{i\theta})  \right) \right\rangle_{\text{disk}} \ ,
\end{equation}
where we take the disk to have Neumann boundary conditions.  As a first step, we separate $X = x_0+\widetilde{X}$ giving
\begin{equation}
    \left\langle \exp\left(-\frac{u}{4}  x_0^2(e^{i\theta})
     -x_0 \frac{u}{4\pi} \widetilde{X}(e^{i\theta}) -\frac{u}{8\pi} \int d\theta :\widetilde{X}^2:(e^{i\theta})   \right) \right\rangle_{\text{disk}} \ .
\end{equation}
Performing the zero mode integral over $x_0$ gives
\begin{equation} \label{Corr}
  \sqrt{\frac{4\pi}{u}} \left\langle \exp \left( -\frac{u}{8\pi} \int d\theta :X^2:(e^{i\theta}) + \frac{u}{16\pi^2} \left( \int d\theta X(e^{i\theta}) \right)^2 \right) \right\rangle^0_{\text{disk}} \ ,
\end{equation}
where the superscript on the correlator indicates that the zeromode integral has been performed.  

The greens function on the disk with Neumann boundary conditions is given by ($\alpha' = 2$ here)
\begin{equation} \label{greensfunction}
 G(z_1,z_2) = - \log |z_1-z_2|^2 -\log|1-z_1 \bar{z}_2|^2 + |z_1|^2 + |z_2|^2 \ .
\end{equation}
We define the boundary normal ordering,
\begin{equation}
  :X(e^{i\theta_1})X(e^{i\theta_2}): = X(e^{i\theta_1})X(e^{i\theta_2}) + 2 \log|1-e^{i(\theta_1-\theta_2)}|^2 \ .
\end{equation}
Computing (\ref{Corr}) using (\ref{greensfunction}) to second non-trivial order gives
\begin{equation}
  K(u) = \sqrt{\frac{4\pi}{u}} \left( 1 +\frac{\pi^2 u^2}{12} + \mathcal{O}(u^3) \right) \ .
\end{equation}
Using Witten's expression for the partition function we would expect to find
\begin{equation} \label{WittenZ}
  K_{\text{expected}}(u) = \sqrt{u} \exp(\gamma u) \Gamma(u) \ .
\end{equation}
Expanding this out for small $u$, one finds
\begin{equation}
 K_{\text{expected}}(u) = \frac{1}{\sqrt{u}} \left( 1 + \frac{\pi^2 u^2}{12} + \mathcal{O}(u^3) \right) \ .
\end{equation}
We see that, up to a constant that is independent of $u$, the two expressions are the same.  Noting that changing our definition of normal ordering of $:X^2:$ would introduce a linear term in $u$ and spoil the equality, we see that the standard normal ordering scheme for $:X^2:$ produces the same result as the direct insertion of $X^2$ into the boundary interaction.

\bibliography{c}\bibliographystyle{utphys}

\end{document}